\documentclass[aps,prb,amsmath,amssymb,twocolumn,superscriptaddress]{revtex4}
\usepackage{graphicx}
\usepackage{braket}
\usepackage{color}
\usepackage{hyperref}
\usepackage[normalem]{ulem}

\newcommand{\uud}{\uparrow\uparrow\downarrow}
\newcommand{\udu}{\uparrow\downarrow\uparrow}
\newcommand{\duu}{\downarrow\uparrow\uparrow}
\newcommand{\uuu}{\uparrow\uparrow\uparrow}
\newcommand{\ddd}{\downarrow\downarrow\downarrow}

\definecolor{darkgreen}{rgb}{0,0.6,0}

\begin{document}

\preprint{APS/123-QED}

\title{Spin of a multielectron quantum dot and its interaction with a neighboring electron}

\author{Filip~K.~Malinowski}
\thanks{These authors contributed equally to this work}
\affiliation{Center for Quantum Devices and Station Q Copenhagen, Niels Bohr Institute, University of Copenhagen, 2100 Copenhagen, Denmark}

\author{Frederico~Martins}
\thanks{These authors contributed equally to this work}
\affiliation{Center for Quantum Devices and Station Q Copenhagen, Niels Bohr Institute, University of Copenhagen, 2100 Copenhagen, Denmark}

\author{Thomas~B.~Smith}
\affiliation{Centre for Engineered Quantum Systems, School of Physics, The University of Sydney, Sydney NSW 2006, Australia}

\author{Stephen~D.~Bartlett}
\affiliation{Centre for Engineered Quantum Systems, School of Physics, The University of Sydney, Sydney NSW 2006, Australia}

\author{Andrew~C.~Doherty} 
\affiliation{Centre for Engineered Quantum Systems, School of Physics, The University of Sydney, Sydney NSW 2006, Australia}

\author{Peter~D.~Nissen}
\affiliation{Center for Quantum Devices and Station Q Copenhagen, Niels Bohr Institute, University of Copenhagen, 2100 Copenhagen, Denmark}

\author{Saeed~Fallahi}
\affiliation{Department of Physics and Astronomy, Station Q Purdue, and Birck Nanotechnology Center, Purdue University, West Lafayette, Indiana 47907, USA}

\author{Geoffrey~C.~Gardner}
\affiliation{Department of Physics and Astronomy, Station Q Purdue, and Birck Nanotechnology Center, Purdue University, West Lafayette, Indiana 47907, USA}

\author{Michael~J.~Manfra}
\affiliation{Department of Physics and Astronomy, Station Q Purdue, and Birck Nanotechnology Center, Purdue University, West Lafayette, Indiana 47907, USA}
\affiliation{School of Electrical and Computer Engineering, and School of Materials Engineering, Purdue University, West Lafayette, Indiana 47907, USA}

\author{Charles~M.~Marcus}
\affiliation{Center for Quantum Devices and Station Q Copenhagen, Niels Bohr Institute, University of Copenhagen, 2100 Copenhagen, Denmark}

\author{Ferdinand~Kuemmeth}
\affiliation{Center for Quantum Devices and Station Q Copenhagen, Niels Bohr Institute, University of Copenhagen, 2100 Copenhagen, Denmark}

\date{\today}

\begin{abstract}
We investigate the spin of a multielectron GaAs quantum dot in a sequence of nine charge occupancies, by exchange coupling the multielectron dot to a neighboring two-electron double quantum dot. For all nine occupancies, we make use of a leakage spectroscopy technique to reconstruct the spectrum of spin states in the vicinity of the interdot charge transition between a single- and a multielectron quantum dot. In the same regime we also perform time-resolved measurements of coherent exchange oscillations between the single- and multielectron quantum dot. With these measurements, we identify distinct characteristics of the multielectron spin state, depending on whether the dot's occupancy is even or odd.  For three out of four even occupancies we do not observe any exchange interaction with the single quantum dot, indicating a spin-0 ground state.  For the one remaining even occupancy, we observe an exchange interaction that we associate with a spin-1 multielectron quantum dot ground state. For all five of the odd occupancies, we observe an exchange interaction associated with a spin-1/2 ground state. For three of these odd occupancies, we clearly demonstrate that the exchange interaction changes sign in the vicinity of the charge transition. For one of these, the exchange interaction is negative (i.e. triplet-preferring) beyond the interdot charge transition, consistent with the observed spin-1 for the next (even) occupancy. Our experimental results are interpreted through the use of a Hubbard model involving two orbitals of the multielectron quantum dot. Allowing for the spin correlation energy (i.e. including a term favoring Hund's rules) and different tunnel coupling to different orbitals, we qualitatively reproduce the measured exchange profiles for all occupancies.
\end{abstract}

\maketitle

\section{Introduction}

Spins in semiconducting nanostructures offer a wide variety of approaches to quantum computing. These include approaches based on gate-defined single-electron quantum dots realized in GaAs/AlGaAs heterostructures~\cite{Nowack2011,Shulman2012,Gaudreau2011,Studenikin2012,Cao2016,Malinowski2017,Bertrand2016}, Si/SiGe quantum wells~\cite{Maune2012,Kim2014,Eng2015,Kawakami2016,Takeda2016} or in MOS nanodevices~\cite{Veldhorst2014,Maurand2016}, as well as spins localized on crystal defects such as phosphorus donors in silicon~\cite{Pla2012,Muhonen2014}.  Along with this range of material choices, spins trapped in quantum dots offer a myriad of possible qubit encodings, including single-\cite{Veldhorst2014,Kawakami2016,Takeda2016}, double-\cite{Foletti2009,Maune2012,Kim2014,Cerfontaine2016} and triple-dot\cite{Medford2013a,Medford2013,Eng2015} schemes, each with distinct advantages.

In contrast to a large body of experimental work on single qubit devices, there are only a handful of demonstrations of two-qubit entangling operations~\cite{Nowack2011,Shulman2012,Nichol2017,Veldhorst2015,Watson2017}, despite their necessity for quantum computing.  Approaches to two-qubit entangling gates based on direct exchange interaction between neighboring tunnel-coupled quantum dots~\cite{Nowack2011,Veldhorst2015,Watson2017} offer fast, high-fidelity operation~\cite{Martins2016,Reed2016}.  Unfortunately, these approaches require dots that are very closely spaced next to each other, which makes fabrication and cross-coupling between qubits a challenge for multi-qubit systems~\cite{Taylor2005a,Wang2015,Zajac2016}.  In contrast, approaches based on direct charge dipole-dipole interaction can offer longer ranges, but suffer from weak coupling (and thus slow gate times) and comparatively lower fidelities~\cite{Shulman2012,Nichol2017}.  Such dipole-dipole interactions could be mediated by superconducting cavities~\cite{Burkard2006,Liu2014,Viennot2015,Mi2016,Russ2015a,Srinivasa2016}, thereby providing a mechanism to couple over even longer ranges, similar to what is commonly used for superconducting qubits.  However, the small dipole moments and susceptibility to charge noise make it unclear whether these approaches will lead to improvements in gate speed and fidelity for spin qubits.

An attractive alternative that has recently been proposed~\cite{Srinivasa2015,Mehl2014a} and demonstrated~\cite{Baart2017,JB_mediated_exchange} is to base two-qubit coupling on exchange interactions, using an intermediate quantum system as a mediator.
This approach makes use of the high speed associated with exchange processes, without the need to arrange quantum dots in direct contact with each other, and is therefore attractive for current fabrication techniques.
In particular, a mesoscopic multielectron quantum dot~\cite{Folk1996,Kouwenhoven1997,Stewart1997,Folk2001,Alhassid2000,Negative-J} could serve as both coupling mediator and spacer~\cite{JB_mediated_exchange,Croot2017}, providing a pathway for scalability to multi-qubit systems.

To serve as a mediator and spacer, the multielectron quantum dot needs to fulfill several requirements:
\begin{enumerate}
	\item The physical size of the multielectron dot should be such that qubit dots can be spaced by at least a few hundred nanometers. This distance facilitates the fabrication of gate electrodes necessary for qubit control and readout.  A large size may also allow the coupling of multiple qubits to the same mediator.
	\item The ground state spin of the multielectron quantum dot must be well defined, to enable the interaction between qubits without entangling with the mediator~\cite{Hu2001}.  Conceptually, a multielectron quantum dot with a non-degenerate spinless ground state appears to be the most straightforward implementation of such a coupler~\cite{Mehl2014a,Srinivasa2015}.
	\item The level spacing of the multielectron quantum dot and the relevant tunnel couplings must be larger than both the energy of the thermal fluctuations ($k_BT\approx 10$~$\mu$eV for $T=100$~mK) and the excitation spectrum of the control voltage pulses ($\approx 20$~$\mu$eV for 5~GHz bandwidth). This condition is necessary to guarantee that the mediator will be prepared in the ground state and to avoid its accidental excitation.
	\item  The ground state spin, level spacing and tunnel coupling of the multielectron quantum dot must be tunable with high yield.  These parameters depend on mesoscopic details of the multielectron dot, and hence cannot be easily controlled by the choice of geometry alone.
	\item The strength of the exchange interaction must provide a competitive timescale for two-qubit gates. Taking 100~ns as an upper target for viable two-qubit gate, this puts a lower bound on the coupling strength of roughly $0.01$~$\mu$eV.
\end{enumerate} 

In this article, we demonstrate that these requirements can be fulfilled by a multielectron quantum dot (except the final requirement,  which we address elsewhere~\cite{JB_mediated_exchange}).  To do this we investigate a linear array of quantum dots in GaAs and configure gate voltages such that an elongated multielectron quantum dot  is populated right next to a two-electron double quantum dot [Fig.~\ref{fig1}(a,b)]. Our approach is based on the fact that the two-electron spin state of the double dot, which can readily be prepared in a singlet state, is sensitive to any spin exchange processes with the neighboring multielectron dot. By pulsing gate voltages towards the charge transition between the right well of the double dot (also referred to as the middle dot) and the multielectron dot, we can systematically induce such spin exchange processes and detect them by subsequent single-shot readout of the double dot. (One may view the double dot as a singlet-triplet qubit, and the presence of spin exchange processes with the multielectron dot as leakage out of the qubit space.) In this way, the double quantum dot serves as a spin-sensitive probe of the multielectron quantum dot.

By employing this double-dot spin probe technique, we study the properties of the multielectron dot in nine subsequent charge occupancies. We are able to identify even and odd occupancy of the dot, and find the following sequence for the ground state spin of the multielectron dot:
With increasing occupancy of the dot, the ground states form a sequence of alternating spin-0 (even occupancy) and spin-1/2 (odd occupancy) states, interrupted once by a spin-1 ground state for a particular even occupancy. 

Moreover, we discover a peculiar behavior of the exchange interaction at the charge transition for the case of spin-1/2 multielectron dot ground state. Namely, the exchange interaction changes sign when changing dot-defining gate voltages by only a few millivolt. A Hubbard model that includes two orbitals of the multielectron dot as well as a triplet-preferring spin correlation energy enables us to reproduce the energetics associated with the total spin of the multielectron quantum dot. From that model we derive a ``phase diagram'' that reveals four regimes with qualitatively distinct energy spectra and associated exchange interaction dependencies.

The article is organized as follows. In Sec.~\ref{sec:setup} we describe in detail the studied sample and the sequences of voltage pulses used to induce interactions between the probe electron in the middle dot and the multielectron quantum dot. In Sec.~\ref{sec:progression} we present the observed sequence of ground states as the occupancy of the multielectron quantum dot is increased one electron at a time. Based on this phenomenology we propose a Hubbard model for the description of the multielectron quantum dot. In Sec.~\ref{sec:0} we present the experimental evidence for a spin-0 ground state for three of the studied electron occupancies. Sec.~\ref{sec:1/2} contains an in-depth study of the interaction between the probe electron and a spin-1/2 state of the multielectron quantum dot, for five different electron occupancies. In Sec.~\ref{sec:1} we present data supporting the observation of a spin-1 ground state. Finally, in Sec.~\ref{sec:summary}, we summarize our results.

\section{Experimental setup and techniques}
\label{sec:setup}

\begin{figure*}[tb]
	\centering
	\includegraphics[width=\textwidth]{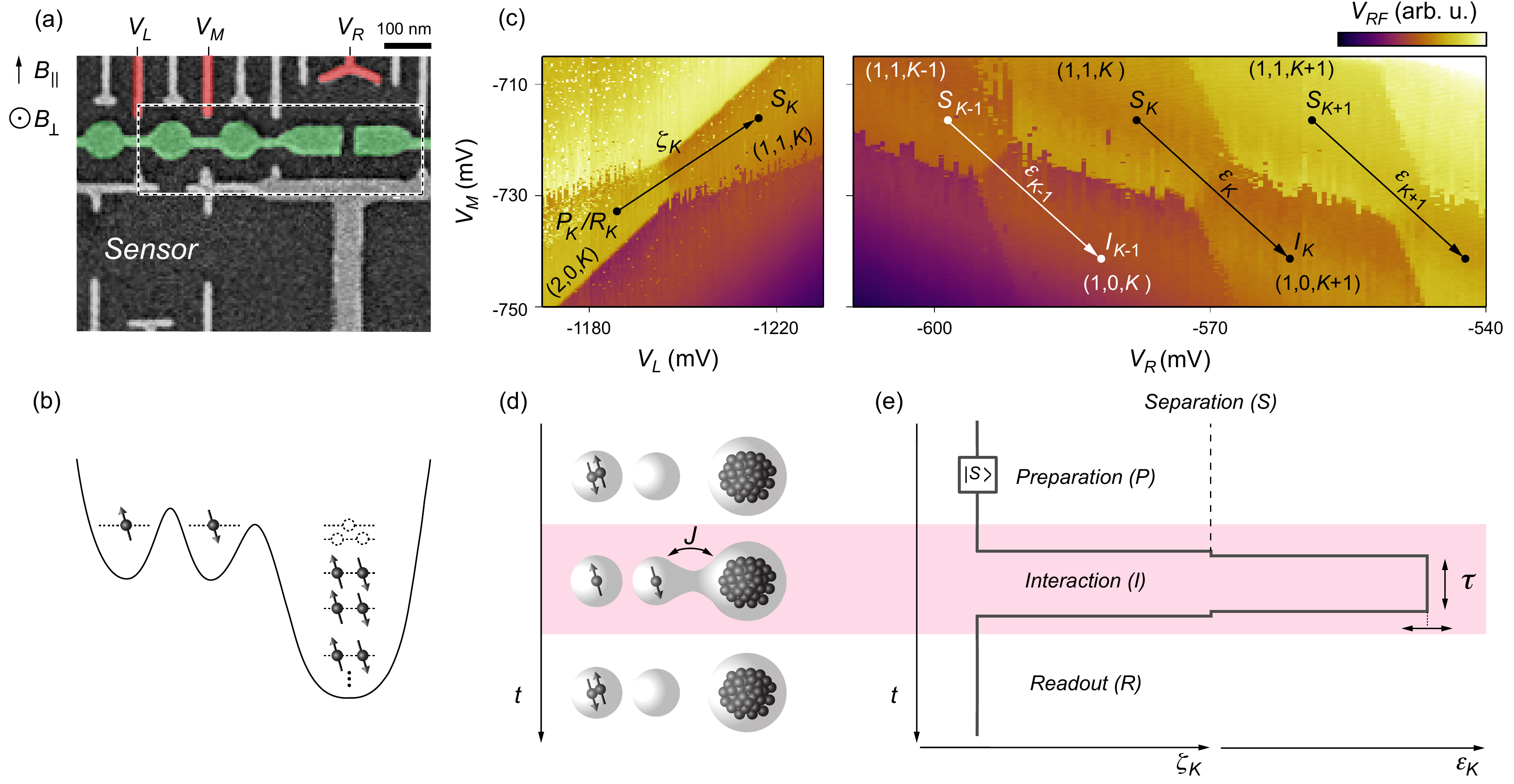}
	\caption{
	(a) Scanning-electron micrograph of the device, colored in light-gray and red to indicate metallic gate electrodes that deplete the 2DEG below the surface (negative voltages). Accumulation gates (colored in green, positive voltages) steepen the resulting confining potential of the quantum dots that form underneath. Voltage pulses applied to the gates $V_{L,M,R}$ control individual electrons in the triple-dot array on a nanosecond timescale.
	(b) Illustration of the electron configuration in the resulting triple quantum dot. We refer to the left and middle dot as the double dot, and to the right dot as the multielectron dot. The precise single-particle level structure and occupation number of the right dot determine the spin properties of the multielectron dot, and are the focus of this experiment. 
	(c) Charge diagrams of the triple quantum dot in the absence of voltage pulses. The left panel shows the interdot charge transition of the two-electron double quantum dot. The right panel presents the charge transition at which the middle electron transfers to the multielectron quantum dot, for different initial occupations of the multielectron dot ($K-1$, $K$, and $K+1$) depending on $V_R$. 
	Labels $P_K$, $R_K$ and $S_K$ indicate positions in gate voltage space at which the electron pair is, respectively, prepared, read out and separated, if appropriate time-dependent voltage pulses are applied. In particular, arrows labeled $\zeta_K$ and $\varepsilon_K$ indicate axes in gate-voltage space used to define the voltage pulses in this experiment. For example, gate voltages associated with the interaction step ($I_K$) are varied systematically (see section~\ref{sec:setup}), but always remain on the $\zeta_K$ or $\varepsilon_K$ axis.
	(d) Operating principle of probing the multielectron spin state by the two-electron double dot. First, a pair of electrons is prepared in a singlet state on the left dot. Next, one of these electrons is transferred to the middle dot, allowing a spin-sensitive interaction ($J$) with the multielectron dot. In the last step, the spin of the middle electron is measured relative to the reference electron in the left dot by means of Pauli blockade. 
	(e) Implementation of (d) by voltage pulses along $\zeta_K$ and $\varepsilon_K$. The outcome of each interaction cycle depends on pulse amplitude ($\varepsilon$) and duration ($\tau$), and, crucially, on the occupation and spin of the multielectron dot.
	}
	\label{fig1}
\end{figure*}

The quantum dots are defined in a GaAs/AlGaAs two-dimensional electron gas (2DEG), with electron density $2.5 \times 10^{15}$ m$^{-2}$ and mobility $230$ m$^2$/Vs. The 2DEG is located 57 nm below the heterostructure surface. A layer of HfO$_2$ with 10~nm thickness is deposited on top of the heterostructure, followed by the patterning of gold electrodes by electron-beam and lift-off lithography. The oxide layer has a double purpose: first, it allows the application of negative and positive gate voltages without resulting in large leakage currents that would appear through the Schottky barrier at the GaAs surface; second, it blocks even minute tunneling events between the gate electrodes and the donor layer in the GaAs heterostructure, which would cause effective charge noise and sample switching behavior~\cite{Buizert2008}. The experiment is performed in a dilution refrigerator with the mixing chamber at 20~mK.

A scanning-electron micrograph of the active part of the device is presented in Fig.~\ref{fig1}(a). The light gray and colored structures are metallic gates that are used to define the quantum dot confining potential. The green-colored accumulation gate is operated at small positive voltage of +40~mV. The remaining gates are operated at negative voltages to deplete the 2DEG and to tune the device. The accumulation gate in this design was introduced to increase the quantum dot potential depth and to improve the tunability of the device.  The resulting distance between single-electron dots, nominally 150~nm center-to-center, is approximately 30\% smaller than in typical designs without the accumulation gate~\cite{Medford2013,Yoneda2014,Nichol2017}.

Under the accumulation gate (indicated in Fig.~\ref{fig1}(a) by a dashed rectangle) two single-electron quantum dots next to a multielectron quantum dot are tuned up, as schematically indicated in~Fig.~\ref{fig1}(b).
Based on the 2DEG density and the device geometry (dot size roughly 120$\times$250~nm) we estimate the electron occupancy of the multielectron quantum dot to be between 50 and 100. The narrow gap in the accumulation gate allows the application of different voltages to different parts of the accumulation gate, but in this study we apply the same (positive) voltage to both parts. A selected number of depletion gates, labeled $V_{L,M,R}$ and shaded in red, are connected to high-bandwidth coaxial lines in the dilution refrigerator, with an associated RC time constant of 0.8~ns. 
To perform sub-microsecond charge and spin manipulations, time-varying voltage pulses are applied to these gates, while DC voltages (applied to all gates) can be modified slowly to explore various occupancies and tunings of the multielectron dot.

In Fig.~\ref{fig1}(c) we present typical charge stability diagrams of the double-dot  multielectron-dot system. On the left we present the charge diagram with respect to gate voltages $V_L$ and $V_M$. These gate voltages are dedicated to control the state of the double quantum dot.  Indeed, this diagram reveals the interdot charge transition of the double dot, occurring between the $(2,0,K)$ and $(1,1,K)$ regions. Here, ($L$,$M$,$R$) indicates the number of electrons in the left, middle and multielectron dot, respectively, and $K$ is an unknown but fixed integer between 50 and 100, which we can vary by changing the DC tuning of the device. By adjusting the voltages $V_L$ and $V_M$ in such a way that these gate voltages follow the detuning axis $\zeta_K$ [Fig.~\ref{fig1}(c), left panel], the charge configuration of the two electrons within the double dot can be controlled without affecting the number of electrons on the multielectron dot.

On the $\zeta_K$ axis we define point $P_K$ ($R_K$) that serves as the preparation (readout) point of the double-dot spin state. We also define a separation point $S_K$ at which the two electrons within the double dot do not interact via exchange with the multielectron quantum dot and only very weakly interact via exchange with each other. 
Having chosen the separation point $S_K$ we can map out the charge stability diagram of the multielectron dot as a function of voltages $V_M$ and $V_R$, as illustrated in the right panel of Fig.~\ref{fig1}(c). In this charge diagram we identify the point $S_K$ in the gate voltage space, and use it to define the $\varepsilon_K$ axis that runs from $S_K$ through the interdot charge transition between the $(1,1,K)$ and $(1,0,K \! + \! 1)$ regions. By controlling the position of gate voltages along this axis (i.e. point $I_K$) we can induce the interaction between the single electron in the middle dot and the multielectron dot, while preserving the reference electronic spin in the left dot. By slightly changing the DC tuning of the quantum dots (in particular $V_R$), we can change the occupancy of the multielectron dot one by one, and define analogous control axes for different charge states. 
In Fig.~\ref{fig1}(c) these are schematically illustrated by axes labeled $\varepsilon_{K-1}$ and $\varepsilon_{K+1}$.

Having defined the points $P_K/R_K$, $S_K$ and detuning axes $\zeta_K$ and $\varepsilon_K$ for each occupancy $K$ of the multielectron quantum dot, we can apply gate voltage pulses that quickly change the charge configuration from $(2,0,K)$ to $(1,1,K)$ to $(1,0,K \! + \! 1)$ and back, thereby allowing the study of interactions between the middle electron with the multielectron quantum dot [illustrated in Fig.~\ref{fig1}(d,e)].  Specifically, the first pulse initiates the system at point $P_K$, resulting in a pair of electrons prepared in the singlet state $\ket{S}$ on the leftmost quantum dot. From there, a pulse to point $S_K$ separates these two entangled electrons while maintaining their spin singlet correlation.  At point $S_K$ we pause for one clock cycle of the waveform generator, which varies between 0.83 and 2.5~ns in this study. This precaution ensures that we indeed transfer the electron through the middle dot to the multielectron dot, instead of ejecting it into one lead followed by injection of another electron to the multielectron dot from the other lead. Because one clock cycle is shorter than the dephasing time due to interaction with the nuclear spins~\cite{Petta2005,Reilly2008,Bluhm2010}, $T_2^* \approx 10$~ns, this waiting time does not significantly affect the singlet correlation of the two electrons.  The next step of the pulse cycle jumps to a point along the $\varepsilon_K$ axis and remains there for time $\tau$.   
It is during this stage that the interaction between the electron and multielectron quantum dot occurs. We then return to $S_K$ for another clock cycle of the waveform generator.  Finally, we pulse back to the $(2,0,K)$ charge configuration at point $P_K/R_K$. The system reaches this charge configuration only if the pair of electrons on the double quantum dot forms a spin singlet state, otherwise the system is Pauli-blocked and remains in the metastable $(1,1,K)$ charge state. The reflectometry readout of the conductance through the neighboring sensor dot allows us to distinguish these charge states, thereby yielding a single-shot spin readout (``singlet'' or ``triplet'') for a reflectometry integration time between 5 and 20~$\mu$s.
Both parameters $\varepsilon_K$ and $\tau$ are varied within a sequence of pulse cycles. 
Using this pulsed-gate technique, coherent spin dynamics and incoherent spin mixing can both be detected, by choosing $\tau$ sufficiently short or long, respectively.

\section{The multielectron quantum dot}
\label{sec:progression}

In this section we present and discuss a theoretical model  that describes the multielectron quantum dot and its tunnel coupling to the double quantum dot. It will be used in subsequent sections to understand the experiments that systematically induce exchange interactions between the single electron residing in the middle dot and the multielectron quantum dot.

Appropriate for a semiconductor with negligible spin-orbit coupling (but allowing for non-trivial electron-electron correlations), we model the multielectron quantum dot with the Hamiltonian:
\begin{equation}
	\hat{H}_R = U_R \hat{n}_R^2 + \sum\limits_{\substack{\lambda \in \mathbb{N} \\ \alpha=\uparrow,\downarrow}} \varepsilon_\lambda \hat{c}_{\lambda,\alpha}^\dag \hat{c}_{\lambda,\alpha} - \frac{\xi}{2} \hat{S}^2,
	\label{eq:JB}
\end{equation}
where $U_R$ is the dot charging energy; $\hat{n}_R$ is the operator counting the total number of electrons; $\hat{c}_{\lambda,\alpha}^{(\dag)}$ are the annihilation (creation) operators for an electron on the single particle level $\lambda$ with spin $\alpha$; $\varepsilon_\lambda$ are the energies of the single particle levels; $\hat{S}$ is the total spin operator and $\xi$ is the spin correlation energy. The subscript $R$ in this formula refers to the multielectron dot as the right dot, whereas $L$ and $M$ denote the left and middle single-electron dots.

\begin{table}[tb]
	\caption{ Summary of the inferred ground state spin for 9 subsequent charge occupancies of the multielectron quantum dot. A sequence of alternating spin-0 and spin-1/2 states is interrupted once by a spin-1 ground state. 
To emphasize the role of electron parity, we have arbitrarily chosen one even dot occupation as a reference, labelled $2N$, and specify other dot occupations relative to that occupation.  
}
	\label{tab1}
\begin{ruledtabular}
\begin{tabular}{lll}
\textbf{Multielectron dot} & \textbf{Inferred ground} & \textbf{Experimental} \\ 
\textbf{occupancy} & \textbf{state spin} & \textbf{evidence} \\ \hline
$2N \! - \! 5$ & 1/2 & Fig.~\ref{fig10} \\ 
$2N \! - \! 4$ & 0 & Fig.~\ref{fig4}(a) \\ 
$2N \! - \! 3$ & 1/2 & Fig.~\ref{fig7} \\ 
$2N \! - \! 2$ & 0 & Fig.~\ref{fig4}(b) \\ 
$2N \! - \! 1$ & 1/2 & Fig.~\ref{fig8} \\ 
$2N$ & 0 & Fig.~\ref{fig3} \\ 
$2N \! + \! 1$ & 1/2 & Fig.~\ref{fig9} \\ 
$2N \! + \! 2$ & 1 & Fig.~\ref{fig13} \\ 
$2N \! + \! 3$ & 1/2 & Fig.~\ref{fig11} \\ 
\end{tabular} 
\end{ruledtabular}
\end{table}

The relative strength of the three terms present in this Hamiltonian determines the spin properties of the multielectron quantum dot. The charging energy of the multielectron quantum dot, $U_R \approx 1$~meV, is estimated from the distance between the multielectron dot charge transitions [$\Delta V_R \approx 20$~mV; Fig.~\ref{fig1}(c)] and the typical lever arm between the gates and the dots in devices of similar design ($\approx 0.05~e$). The charging energy may vary slightly as a function of the dot occupancy, as additional electrons may increase the effective size of the quantum dot (soft confining potential). For the results presented here it is only relevant that the charging energy is much larger than the other energy scales, discussed below.  

From the lithographic size of the device we estimate the typical level spacing~\cite{Datta1997} to be $\langle \Delta E \rangle = \pi \hbar^2/m^* A \approx 0.12$~meV, where $\hbar$ is the reduced Planck constant, $m^*$ is the effective electron mass in GaAs and $A$ is the area of the 2-dimensional quantum dot. However, the lack of symmetry causes the level spacings to vary. 
The determination of level spacings $\Delta E$ and correlations between them for a particular mesoscopic quantum dot is a formidable theoretical task. Their distributions are typically described using random matrix theory with the orthogonal ensemble~\cite{Folk1996,Folk2001,Brouwer1999,Kurland2000,Baranger2000}, which by itself neglects interaction effects. 
Interaction effects can be introduced by means of random-phase approximation~\cite{Blanter1997}, mean-field approximation\cite{Baranger2000}, density functional theory~\cite{Hirose2002}, the Anderson model~\cite{Sivan1996} or by an on-site Hubbard interaction term~\cite{Brouwer1999} (for review see Ref.~\onlinecite{Alhassid2000}). For the results presented here we consider the case where the width of the level spacings distribution, $\sigma_{\Delta E}$, is comparable to $\Delta E$ and $\xi$, allowing for the emergence of mesoscopic magnetism~\cite{Gorokhov2004}. Also, we assume that the single particle energies do not depend on the occupancy of the dot, consistent with earlier experiments that found that the excitation spectra of (few) subsequent charge states are highly correlated~\cite{Stewart1997}.

The spin correlation energy $\xi$ is the most difficult quantity to estimate due to the lack of data in the literature.  We make the assumption that $\xi/2$ is comparable, but smaller than, $\langle \Delta E\rangle$, based on two experimental observations. First, we observe no significant polarization of the electronic spins, which would be expected from the Stoner instability for the case\cite{Andreev1998,Kurland2000,Alhassid2000} $\xi/2 > \langle \Delta E\rangle$. Second, we observe the occurrence of ground states with spin $>1/2$, which indicates that the single particle energy of the first excited state sometimes becomes smaller than $\xi$. This excludes the possibility $\xi \ll \langle \Delta E\rangle$. Throughout the paper we will use $U_R=1$~meV and $\xi=0.1$~meV, while we will use different $\Delta E$ for different occupancies of the multielectron quantum dot (justified by level filling of a mesoscopic spectrum).

Our study involves modeling of the interaction between the spin occupying the middle dot $M$ [Fig.~\ref{fig1}(b)] and one or two lowest empty or partially occupied orbitals of the multielectron quantum dot. The occupancy of the multielectron quantum dot will determine the nature of this interaction.  We consider three cases, ordered by increasing complexity:
\begin{itemize}
	\item All levels of the multielectron quantum dot are either empty or doubly occupied, and the total spin is zero.  In this case, the interaction of the double quantum dot with the multielectron quantum dot can be modeled as an effective interaction of the spin of the middle dot $M$ tunnel coupled to one unoccupied orbital in the right dot, $R$ (Sec.~\ref{sec:0}).
	\item There is exactly one unpaired spin in the multielectron quantum dot, giving rise to a total spin 1/2 (Sec.~\ref{sec:1/2}).  In this case, the interaction of the double quantum dot with the multielectron quantum dot can be modeled as an effective interaction of the spin of the middle dot $M$ tunnel-coupled to a single spin in the right dot $R$.  Depending on the details of the spin interaction terms of the multielectron quantum dot, we must consider both the partially occupied orbital of the multielectron quantum dot as well as the lowest unoccupied orbital.
	\item Several unpaired spins in the multielectron quantum dot form a nonzero total spin, e.g., spin 1 (Sec.~\ref{sec:1}). This requires the modeling of two closely spaced orbitals in the right dot, $R1$ and $R2$, in conjunction with a sufficiently large spin correlation energy $\xi$ relative to the orbitals' spacing $\Delta E$.
\end{itemize}

In our experimental data, to be discussed in detail in sections~\ref{sec:0}-\ref{sec:1}, we find that all three cases do occur as we explore different occupation numbers of the multielectron dot. 
Table~\ref{tab1} provides an overview of the observed sequence of alternating spin-0 and spin-1/2 ground states. 
This sequence is interrupted once by a spin-1 ground state instead of spin-0 (with profound implications for the associated exchange profiles).
The occurrence of a non-minimal ground state spin in our experiments corroborates earlier findings in GaAs quantum dots by Folk \emph{et al.}\cite{Folk2001} and Lindemann \emph{et al.}\cite{Lindemann2002}, who identified ground states spins by studying the change of the Coulomb peak spacing in magnetic field.

\section{Spin-0 behavior for even occupancies  ($K=2N \! - \! 4$, $2N \! - \! 2$, $2N$)}
\label{sec:0}

\begin{figure}
	\centering
	\includegraphics[width=0.48\textwidth]{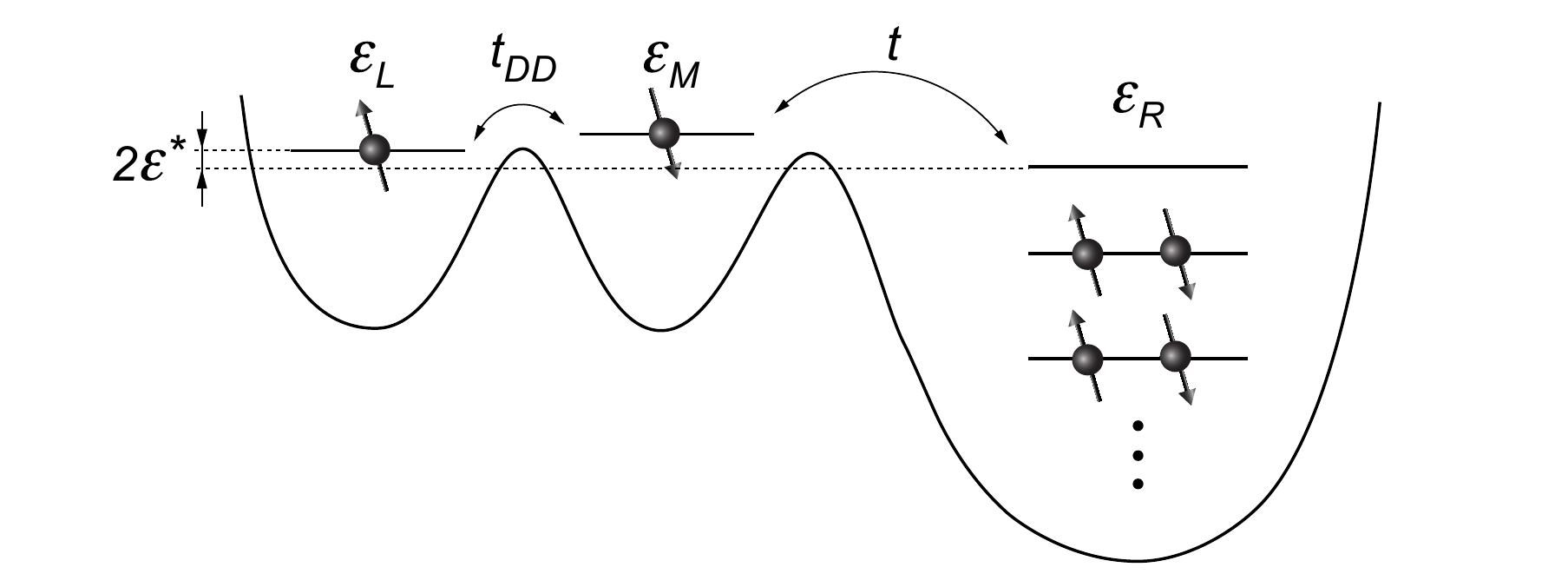}
	\caption{
	Schematic of the even-occupied spinless multielectron dot, coupled to the two-electron double dot. 
	Symbols $\varepsilon_{L/M/R}$ indicate the single-particle energies of the lowest orbitals in the double dot and the lowest unoccupied orbital in the multielectron dot. Arrows indicate tunnel couplings within the double dot ($t_{DD}$) and between the middle and the multielectron dot ($t$). For simulations, the detuning of the right dot relative to the left dot is varied ($\varepsilon^*$), which allows the generation of energy spectra (as in Fig.~\ref{fig3}b) and comparison to experimental  leakage spectroscopy data (see discussion of Figs.~\ref{fig3}c,d).
	}
	\label{fig2}
\end{figure}

\begin{figure}
	\centering
	\includegraphics[width=0.48\textwidth]{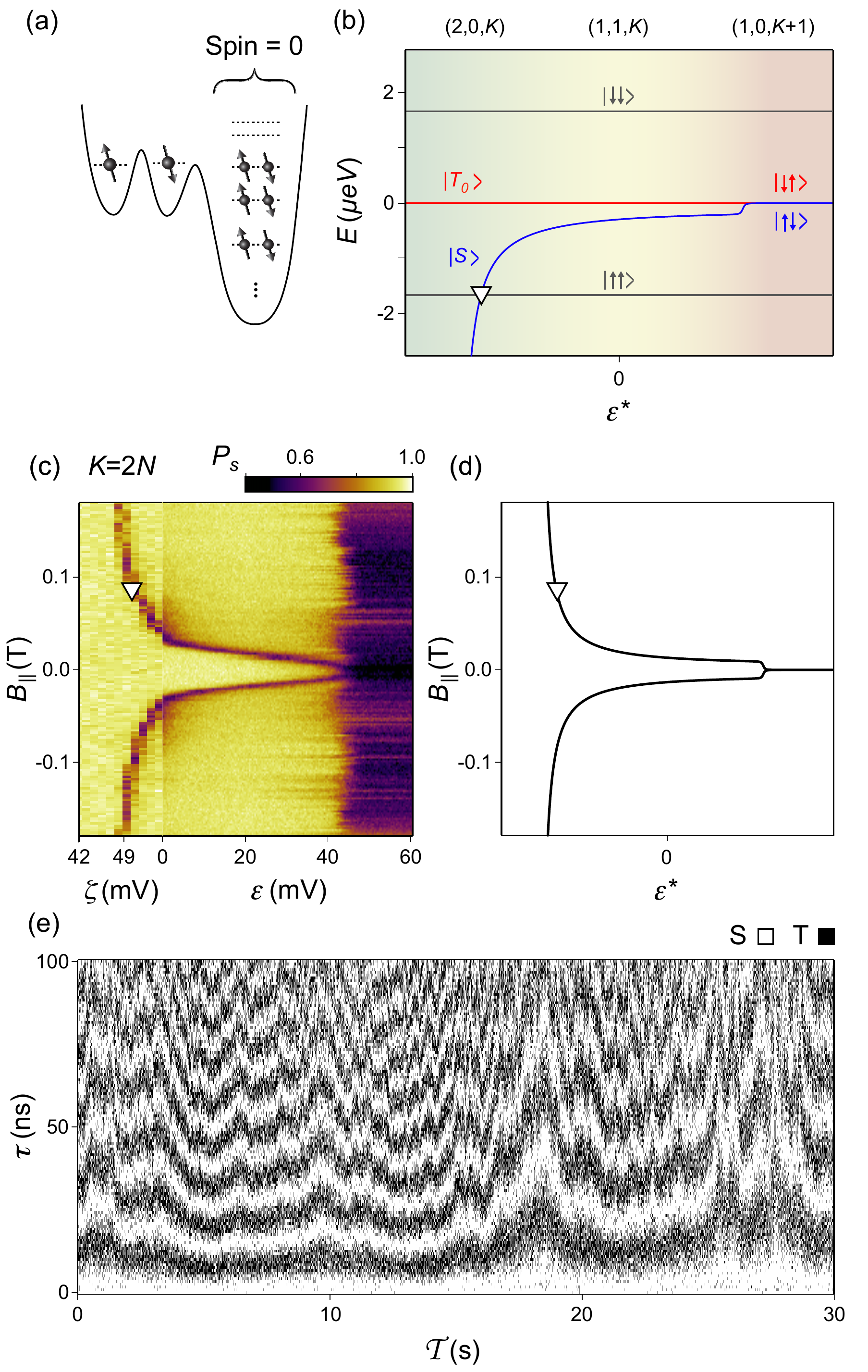}
	\caption{Spin-0 ground-state behavior of the $2N$ occupied multielectron dot. (a) Schematic representation of the electron configuration for even-occupied multielectron dot with spin-0 ground state. Electrons below the Fermi energy in the multielectron dot are assumed to pair up into spin singlets.
	(b) Eigenstates of a two-electron triple dot system for fixed applied magnetic field, calculated as a function of $\varepsilon^*$. For suitable input parameters the resulting spin states (labeled) are a useful model for comparison with experimental data obtained from multielectron dot charge states $(2,0,2N)$, $(1,1,2N)$ and $(1,0,2N \! + \! 1)$ (see main text). 
	(c) Experimental leakage spectroscopy for $K=2N$, revealing strong exchange coupling in $(2,0,2N)$, i.e. within the double dot, and vanishing exchange interaction in $(1,0,2N \! + \! 1)$. We associate the sharp feature of suppressed $P_S$ (white triangle) with the $S$-$T_+$ crossing and the overall suppression of $P_S$ above $\varepsilon \approx 45$~mV with $S$-$T_0$ oscillations arising from Overhauser gradients.
	(d) Leakage spectrum expected from the $S$-$T_+$ crossing within the Hubbard model (white triangle in panel (b)). 
	(e) Time-resolved measurement of coherent oscillations between $\ket{S}$ and $\ket{T_0}$ two-electron spin states in $(1,0,2N \! + \! 1)$ charge state. The observed fluctuations of precession frequencies with laboratory time $\mathcal{T}$ are characteristic of fluctuating Overhauser field gradients arising from nuclear spin dynamics within the GaAs sample.
	}
	\label{fig3}
\end{figure}

We first focus on even occupancies of the multielectron dot, specifically $2N \! - \! 4$, $2N \! - \! 2$ and $2N$, and show experimental evidence that these have a spin-0 ground state.  (Here, $2N$ indicates a specific even number of electrons, estimated to lie between 50 and 100).  
Provided that the spin correlation term is smaller than the level spacing between the ground and the first excited state, the model introduced in Sec.~\ref{sec:progression} suggests that the ground state of the multielectron dot can be thought of as an effective vacuum state, i.e., all single-particle states below the Fermi energy are occupied by singlet pairs of electrons [Fig.~\ref{fig2} and \ref{fig3}(a)].
We therefore expect that the double dot will interact with the multielectron dot as if it was an unoccupied dot, and the spin of an electron tunneling into an unoccupied orbital of the multielectron dot would not experience any exchange dynamics.  In this sense, the double dot coupled to the multielectron dot with even occupancy should be qualitatively similar to a two-electron triple dot.

We describe this situation using a phenomenological model based on the Hamiltonian for the multielectron dot detailed in Sec.~\ref{sec:progression}, augmented by terms for the neighboring tunnel-coupled two-electron double quantum dot. Appropriate for spinless even-occupancy ground states we neglect orbitals of the multielectron dot below the Fermi energy (these are occupied by spin-singlet electron pairs). We also neglect all but the lowest unoccupied orbital, arriving at a Hubbard model for the three dots, each having a single orbital, labeled $L$, $M$, and $R$: 
\begin{align}
	\hat{H}_{\text{spin-0}} &= \sum_{i=L,M,R} \left( \varepsilon_i \hat{n}_i + \frac{U_i}{2}\hat{n}_i(\hat{n}_i-1) \right) + \sum_{i\neq j}\frac{K_{ij}}{2}\hat{n}_i\hat{n}_j \nonumber \\
    &- t_{DD} \sum\limits_{\alpha=\uparrow,\downarrow}  (\hat{c}_{L,\alpha}^\dag \hat{c}_{M,\alpha} + \hat{c}_{M,\alpha}^\dag \hat{c}_{L,\alpha}) \nonumber \\ 
	&- t\sum\limits_{\alpha=\uparrow,\downarrow}  (\hat{c}_{M,\alpha}^\dag \hat{c}_{R,\alpha} + \hat{c}_{R,\alpha}^\dag \hat{c}_{M,\alpha}) \,,
	\label{eq:spin0Ham}
\end{align}
where $\hat{n}_i$ is the operator counting the number of electrons on each dot. As illustrated in Fig.~\ref{fig2} the term $\varepsilon_i$ describes the gate-tunable chemical potential of the left $L$, middle $M$ and the right multielectron $R$ dot. $U_i$ and $K_{ij}$ represent, respectively, on- and off-site Coulomb interaction energies. The second and final line incorporate the tunnel coupling $t_{DD}$ within the double dot and the tunnel coupling $t$ between the middle and right dot.

This effective Hamiltonian can be solved in the 2-electron configuration to yield the energy of all possible spin states, using input parameters motivated by experiment.  
For simplicity we plot  in Fig.~\ref{fig3}(b) the eigenenergies as a function of electrostatic detuning between right and left dot, $\varepsilon^*\equiv(\varepsilon_R-\varepsilon_L)/2$. Qualitatively, the dependence on $\varepsilon^*$ can be compared with the observed dependence on the experimental detuning parameters $\zeta_K$ and $\varepsilon_K$ in Fig.~\ref{fig3}(c). For the calculated energy spectrum in Fig.~\ref{fig3}(b) we use $t = 30$~$\mu$eV, $t_{DD} = 15$~$\mu$eV and  --- for clarity --- a non-zero magnetic field (remaining parameters, fixed throughout the paper, are specified in Appendix~\ref{app:parameters}). 
Recall that the ``unoccupied'' state of the multielectron dot is assumed to be an effective ``vacuum'' state with $2N$ electrons in a spin-0 configuration, i.e., the evolution of charge states from (2,0,$2N$) via (1,1,$2N$) to (1,0,$2N \! + \! 1$) in the experiment should be compared to the evolution from (2,0,0) via (1,1,0) to (1,0,1) in the model [indicated by the colored background shading in Fig.~\ref{fig3}(b)]. 
Specifically, we are interested in the evolution of the singlet double-dot state, $\ket{S}$, and the unpolarized triplet double-dot state, $\ket{T_0}$,  as their splitting is a witness of exchange effects. 
Their calculated dependence on $\varepsilon^*$ in Fig.~\ref{fig3}(b) can be understood as follows. In the (2,0,$2N$) charge state (i.e. towards negative $\varepsilon^*$), the singlet and triplet states of the two-electron double dot are split by the well-known intradot exchange energy. This splitting arises from the Pauli exclusion principle (single-particle level spacing of the left dot, modified by small corrections arising from weak correlation effects), although in our model it diverges because we consider only a single orbital in each dot. Towards the (1,1,$2N$) tuning, i.e. $\varepsilon^*$=0, the exchange splitting gradually decreases as the overlap of the electronic wavefunctions decreases. Finally, in the (1,0,$2N \! + \! 1$) tuning, the exchange splitting reduces to zero, as the two electrons occupy distant dots. 
In this large-$\varepsilon^*$ limit, we label the two (degenerate) states by basis states $\ket{\uparrow \downarrow}$ and $\ket{\downarrow \uparrow}$  (rather than singlet and triplet), as these states are known to become the correct energy eigenstates if differences between the total effective magnetic field in each dot are taken into account (caused for example by uncontrolled Overhauser fluctuations, which are omitted in our effective Hamiltonian).  
In Fig.~\ref{fig3}(b) we also label the fully polarized spin states as $\ket{\uparrow \uparrow}$ and  $\ket{\downarrow \downarrow}$. 
For negative $\varepsilon^*$ they correspond to the well-known fully polarized triplet states within a two-electron double-dot system.

Using spin leakage spectroscopy (which can be viewed as an extension of the ``spin funnel'' measurement of a double quantum dot~\cite{Petta2005,Maune2012}) we experimentally map out the exchange profile across the three charge configurations $(2,0,K)$, $(1,1,K)$ and $(1,0,K \! + \! 1)$. 
By applying the pulse sequence introduced in section \ref{sec:setup}, we prepare the double dot in a singlet state $\ket{S}$, and then pulse to the various interaction points $I_{K}$ along the $\zeta_{K}$ and $\varepsilon_{K}$ axes using a fixed interaction time, $\tau$. 
By choosing $\tau$ sufficiently long (here 150~ns) incoherent mixing between the middle spin and other spin states can be detected, as any such processes reduce the probability, $P_S$, of detecting a spin-singlet state when pulsing back to the readout configuration of the double dot. 
We repeat this procedure for various values of the in-plane magnetic field, up to $B_\parallel= 200$~mT, and associate any significant decrease in $P_S$ with leakage from the singlet state.

\begin{figure}[tb]
	\centering
	\includegraphics[width=0.48\textwidth]{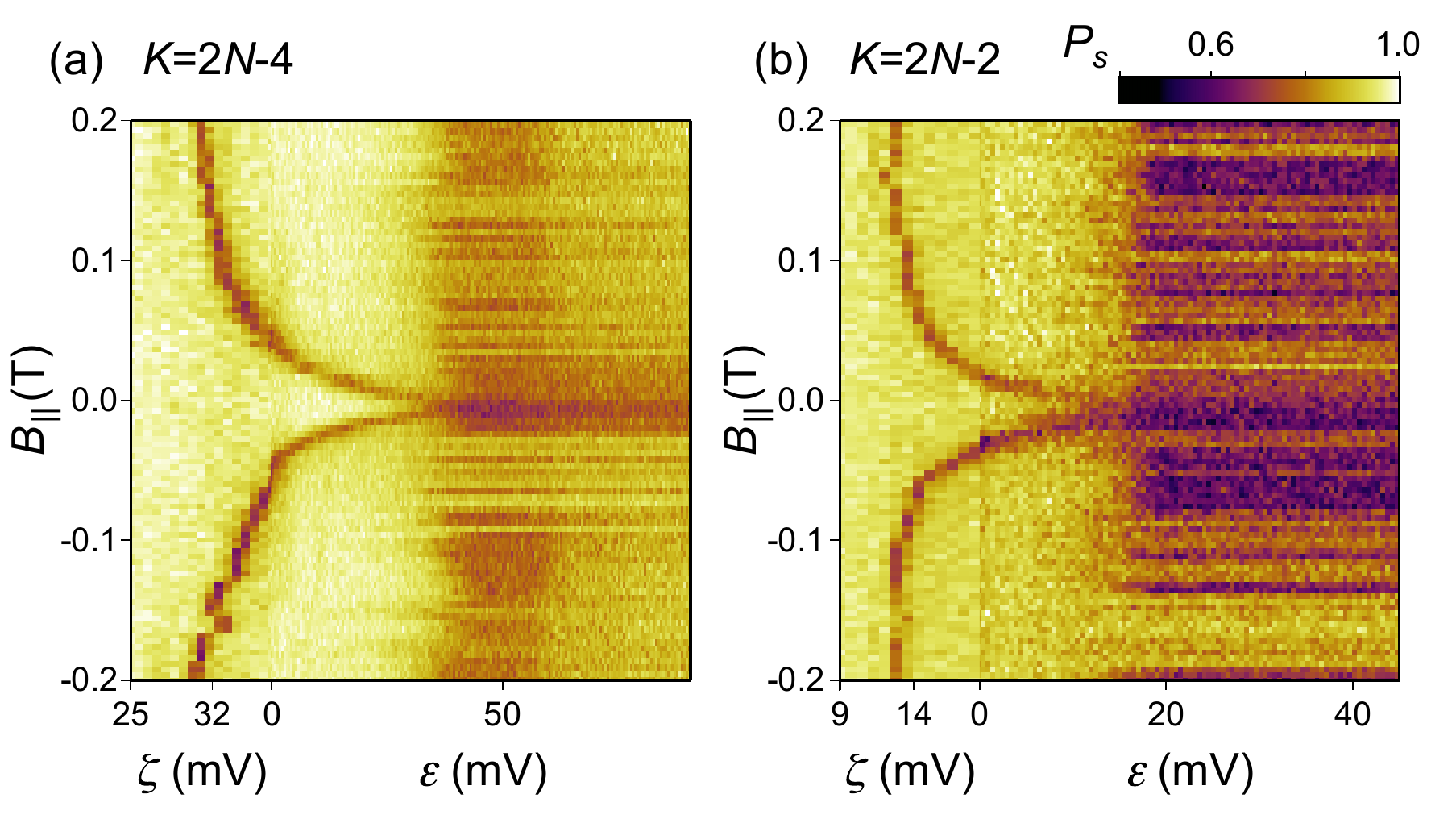}
	\caption{
	Leakage spectroscopy of the double dot coupled to the multielectron dot at the transition between 
	(a) $(2,0,2N \! - \! 2)$, $(1,1,2N \! - \! 2)$ and $(1,0,2N \! - \! 1)$ and
	(b) $(2,0,2N \! - \! 4)$, $(1,1,2N \! - \! 4)$ and $(1,0,2N \! - \! 3)$.
	}
	\label{fig4}
\end{figure}

The result is shown in Fig.~\ref{fig3}(c) for one particular even occupation of the multielectron dot, $K=2N$.
Clearly, there is a sharp feature of reduced $P_S$, marked by a white triangle, that depends on the applied magnetic field.  
We associate it with the crossing of the singlet state $\ket{S}$ and the fully polarized triplet state $\ket{T_+} = \ket{\uparrow\uparrow}$. 
(At negative magnetic field, mixing between $\ket{S}$ and $\ket{T_-} = \ket{\downarrow\downarrow}$ causes an analogues feature, leading to a leakage spectrum that is symmetric with respect to $B_\parallel=0$.)
Indeed, at such crossings rapid mixing due to uncontrolled Overhauser gradients is expected to occur, changing electronic spin projections by 1 on a timescale of $T_2^*\approx 10$~ns \cite{Petta2005}. This leakage feature diverges to high field in the (2,0,$2N$) configuration, indicating that the exchange interaction between the two electrons within the double dot and the single particle spacing in the left dot are relatively large. (Here, we have used that for this particular crossing the associated external magnetic field $B_\parallel$ can be converted into energy using the Zeeman shift associated with $\ket{\uparrow\uparrow}$, i.e. $g\mu_B |B_\parallel|$, where $g \approx 0.4$ is the electronic g-factor for GaAs and $\mu_B$ is the Bohr magneton). 
Towards the (1,1,2$N$) configuration the leakage feature gradually moves towards $B_\parallel = 0$, indicating a decrease of the exchange interaction strength. Finally, it converges to zero field in the (1,0,$2N \! + \! 1$) configuration, consistent with the two electrons being spatially separated and no longer exchange coupled. In this configuration, we also observe a decreased singlet return probability that is independent of the applied magnetic field and $\varepsilon$. We associate this decrease with the mixing between $\ket{S}$ and the unpolarized triplet state, $\ket{T_0}$, driven by the Overhauser field gradient between the left and the multielectron dot. Similar features for the $2N \! - \! 4$ and $2N \! - \! 2$ occupation of the multielectron dot are presented in Fig.~\ref{fig4}, and are reminiscent of analogues $S$-$T_0$ mixing in two-electron double dots with sufficiently small exchange coupling~\cite{Petta2005}.

The observed leakage spectrum in Fig.~\ref{fig3}(c) can be reproduced qualitatively from our Hubbard model, even though it does not take into account any Overhauser gradients. First, we calculate the energy spectrum of the the Hamiltonian of Eq.~\ref{eq:spin0Ham} using the same parameters as above, but with the addition of a Zeeman energy due to the applied in-plane magnetic field, $B_\parallel$. We then identify the ground state associated with preparation of the double dot in the singlet state, and plot those values of $\varepsilon^*$ and $B_\parallel$ for which this state crosses the fully polarized state $\ket{T_+}$. 
(For $B_\parallel<0$ we plot the crossing of the ground state with $\ket{T_-}$.)
The state crossing in our model indicates where spin mixing due to Overhauser gradients is expected [Fig.~\ref{fig3}(d)], in qualitative agreement with experimental data [Fig.~\ref{fig3}(c)].

To confirm the origin of the decreased singlet probability in the $(1,0,2N \! + \! 1)$ configuration, we perform a time-resolved measurement of the Overhauser field gradient~\cite{Barthel2009,Foletti2009,Barthel2012,Shulman2014,Delbecq2016,Malinowski2017a} between the leftmost dot and the multielectron dot. For that purpose, we fix the interaction point in the $(1,0,2N \! + \! 1)$ charge configuration and cyclically vary the waiting time $\tau$ from 0 to 100 ns. The cycle is repeated continuously, keeping track of both the waiting time ($\tau$) and laboratory time stamp ($\mathcal{T}$) associated with each single-shot readout. 
When plotting all individual single-shot readouts (singlet or triplet) versus their associated $\tau$- and $\mathcal{T}$-values [Fig.~\ref{fig3}(e)], coherent oscillations between the singlet $\ket{S}$ and triplet $\ket{T_0}$ state become apparent within each column, with an oscillation frequency that slowly changes from column to column. This fluctuating behavior (shown here over a 30-second long laboratory time interval) is the hallmark of two-electron spin coherence interacting with the (diffusive) dynamics of the GaAs nuclear spin bath, and has been characterized in detail for two-electron double dots~\cite{Reilly2008,Barthel2009,Malinowski2017a,Delbecq2016}.

\section{Spin-1/2 behavior for odd occupancies}
\label{sec:1/2}

For odd occupation number the multielectron ground state must be spinful (found to be spin-1/2 in all cases we study), and accordingly the resulting coupled spin system is more complex compared to the spinless case discussed in section~\ref{sec:0}.  To put our double-dot spin probe technique into context of previous experiments, we begin this section by reviewing a triple quantum dot in the three-electron regime, before we turn to multielectron effects. 
The three-electron regime, (1,1,1), is difficult to realize in the geometry shown in Fig.~\ref{fig1}(a). Therefore, on the same chip we activated another triple dot in which the lithographic size of the right dot is the same as the left and middle one-electron dot, and present measurements for the (1,1,1) regime of that device. 

After describing the relevant physics of this tunnel-coupled (1,1,1) system, using control parameters $\zeta$ and $\varepsilon$ as introduced in Fig.~\ref{fig1}(c), we present multielectron effects for the $(1,1,K)$ system, where $K$ is large and odd. 
The measurements involve several odd occupancies of the multielectron dot and reveal that associated exchange profiles fall into two characteristically distinct categories. Both categories can be reproduced within our Hubbard model, which additionally predicts two other types of exchange profiles that are not observed in the multielectron device (see Subsection~\ref{subsec:phase_diagram}).

\subsection{Review: Three-electron triple quantum dot  ($K=1$)}
\label{subsec:TQD}

\begin{figure}[tb]
	\centering
	\includegraphics[width=0.48\textwidth]{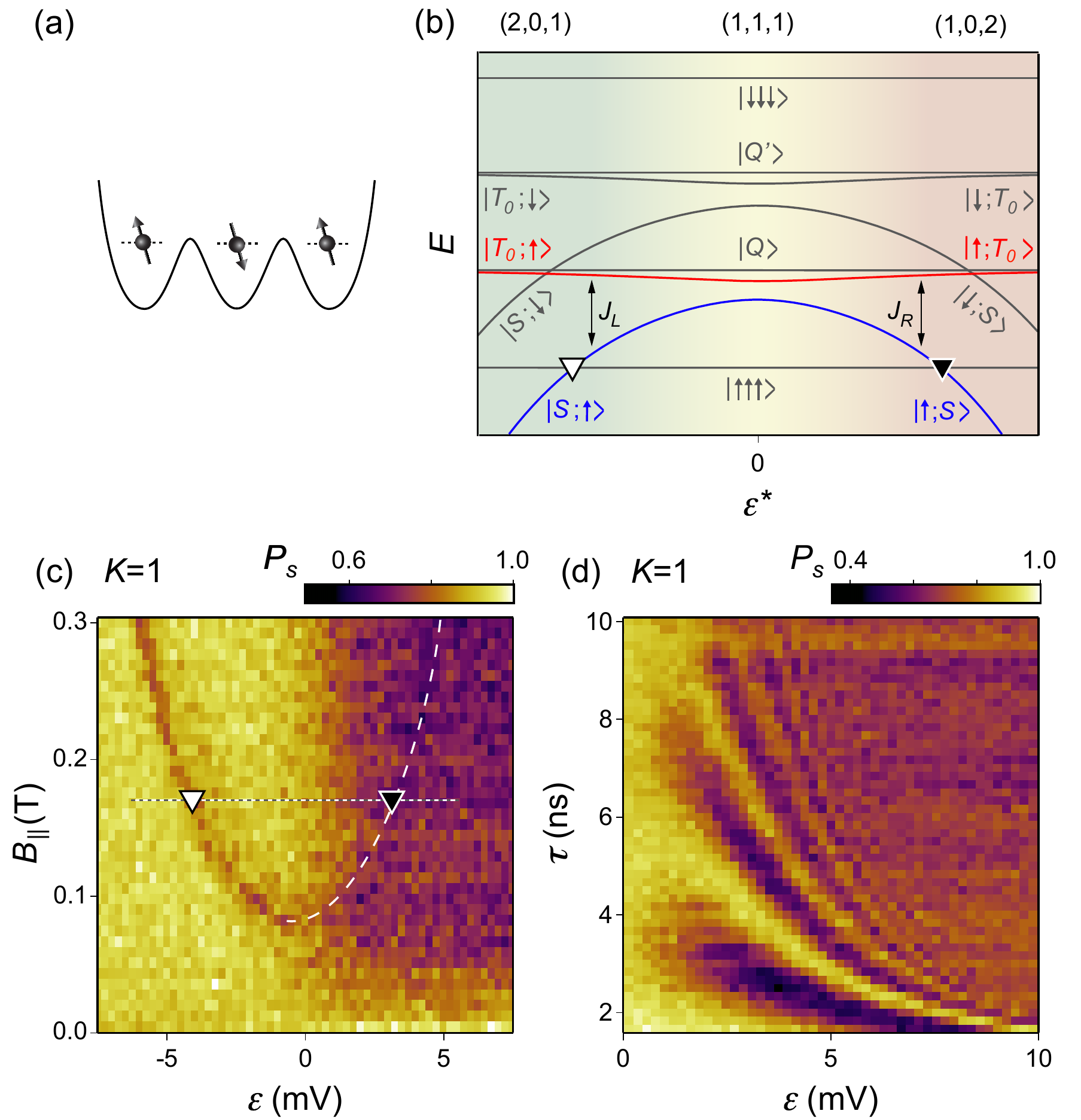}
	\caption{
	(a) Schematic of the three-electron triple dot, which serves as a reference for discussing the odd occupied multielectron dot tunnel-coupled to the double dot.
	(b) Energy diagram of the three-electron triple-dot spin states for a finite external magnetic field. 
	(c) Measured leakage spectrum for the three-electron triplet dot. Markers indicate leakage features attributed to the level crossings marked in (b).
	(d) Measured exchange oscillations reveal a monotonously increasing frequency, corresponding to monotonously increasing $J_R(\varepsilon)$ in (b). 
	}
	\label{fig5}
\end{figure}

The (1,1,1) charge state of a triple quantum dot, schematically illustrated in Fig.~\ref{fig5}(a), allows for $2^3$ distinct spin states. The energy of these 8 three-electron states at finite external magnetic field are shown in Fig.~\ref{fig5}(b). For $\varepsilon^*=0$, all degeneracies are removed by a combination of linear Zeeman coupling (independent of $\varepsilon^*$) and finite interdot tunneling (charge hybridization, $\varepsilon^*$ dependent). 
In particular, the state plotted in blue transforms (smoothly due to interdot tunneling) into a (2,0,1) charge state when reducing $\varepsilon^*$.
Accordingly, the spin state in this limit becomes equivalent to a spin singlet in the left dot, and a spectator spin in the right dot, represented as $\ket{S; \uparrow}$.
Unlike this ``singlet-like state'', the state marked in red displays ``triplet-like'' 
behavior
\footnote{
Conventionally, the terms ``singlet'' and ``triplet'' are used for two-electron spin-singlet and two-electron spin-triplet states, and one may be tempted to use these terms as synonyms for the symmetry under exchange of two spins. Care must be taken when giving in to this temptation for three-electron states, as their total antisymmetry arises in general from a combination of orbital and spin symmetries. In this work, we therefore use ``singlet-like'' and ``triplet-like'' only for states that can be written as a product state of one spectator spin ($\ket{\uparrow}$ or $\ket{\downarrow}$) and a two-electron spin state, which may be a spin-singlet state ($S$) or one of three possible spin-triplet states ($T_+, T_0, T_-$). 
Similarly, four-electron states relevant for Section~\ref{sec:1} behave under certain circumstances like the tensor product of a spin-1 spectator (located in the right dot) and a singlet or triplet two-electron spin state within the left-dot middle-dot double dot.
}: 
due to Pauli blockade it retains its (1,1,1) charge character when reducing $\varepsilon^*$, and smoothly turns into a $T_0$-like state in the left double well and a spectator spin in the rightmost dot, represented as $\ket{T_0; \uparrow}$. 
Many features of the spectrum in Fig.~\ref{fig5}(b) have been studied previously in the context of exchange-only triple-dot spin qubits~\cite{Laird2010,Gaudreau2011,Medford2013a,Poulin-Lamarre2015}. 

Of interest here is the exchange energy between the singlet-like and triplet-like state. In the vicinity of the (2,0,1)-(1,1,1) charge transition we label this energy $J_L$, to indicate that it predominantly arises from tunneling across the left barrier (tunneling across the right barrier is suppressed, as the right dot is in deep Coulomb blockade at this detuning). 
Conversely, in the vicinity of the (1,1,1)-(1,0,2) charge transition, exchange processes across the left barrier are negligible, while $J_R>0$ is significant.  
Accordingly, each eigenstate is labelled by its approximate spin texture, which in this region is a tensor product of the spin in the left dot,  $\ket{\uparrow}$ or  $\ket{\downarrow}$, and the two-electron spin state of the right double quantum dot, $\ket{S}$ or $\ket{T_i}$ (where $i=0,+,-$). 

The splitting between the singlet-like state and the triplet-like state [colored, respectively, blue and red in Fig.~\ref{fig5}(b)], with respect to difference $\varepsilon$ of gate voltages controlling occupancy of the left and right quantum dot, can be mapped out using leakage spectroscopy~\cite{Poulin-Lamarre2015} [Fig.~\ref{fig5}(c)] with a procedure similar to the one employed in Sec.~\ref{sec:0}. In this case the system is prepared in the $\ket{S ; \uparrow}$ state, and the sharp feature of reduced $P_S$ indicates leakage from this singlet-like state to the fully polarized $\ket{\uuu}$ state [white and black triangles in Fig.~\ref{fig5}(b,c)]. We observe that this feature diverges to high magnetic field for large positive and negative values of $\varepsilon$, consistent with the decrease of the energy of the singlet-like state $\ket{S; \uparrow}$ or $\ket{\uparrow; S}$ in the (2,0,1) or (1,0,2) electron configuration, respectively.

As a side note, we mention that four of the $2^3$ triple-dot spin states form a $S=3/2$ quadruplet. An external magnetic field splits these according to spin projections $S_z=\pm 3/2$, labelled as $\ket{\uparrow \uparrow \uparrow}$ and $\ket{\downarrow \downarrow \downarrow}$, and $S_z=\pm 1/2$, labelled as $Q$ and $Q'$. 

The background of the leakage spectrum in Fig.~\ref{fig5}(c) also shows an overall drop of $P_S$ with increasing $\varepsilon$, independent of the applied magnetic field.
This indicates that the eigenstates on the left side ($\varepsilon<0$) differ from the right side ($\varepsilon>0$) of the spectrum. 
Once again, insight can be gained by reducing the interaction time $\tau$, which for $\varepsilon>0$ reveals coherent exchange oscillations between the middle spin and the right dot [Fig.~\ref{fig5}(d)]. 
The frequency of the oscillations increases for larger values of $\varepsilon$, quantifiying the increasing exchange coupling $J_R$ between the middle and the right quantum dot. This precession was previously exploited for the operation of the exchange only qubit~\cite{Laird2010,Medford2013,Eng2015}. For this article, it serves the purpose of exemplifying that the spin of the right dot can be probed coherently using a proximal two-electron double dot. 

This concludes our review of the three-electron triple dot. In the following sections we will extend the same experimental concepts, namely leakage spectroscopy and measurement of exchange oscillations, to the system consisting of the two-electron double quantum dot coupled to the multielectron dot with an odd-occupancy spin-1/2 ground state. The role of $J_L$ and $J_R$ will be played, respectively, by exchange coupling within the double dot, $J_{DD}$, and exchange coupling between the middle spin and the multielectron dot, $J$ (compare Fig. \ref{fig5}(b) and \ref{fig7}(b) discussed below).

\subsection{Negative exchange interaction at the charge transition ($K=2N \! - \! 3$, $2N \! - \! 1$)}
\label{subsec:positive-J}

\begin{figure}[tb]
	\centering
	\includegraphics[width=0.48\textwidth]{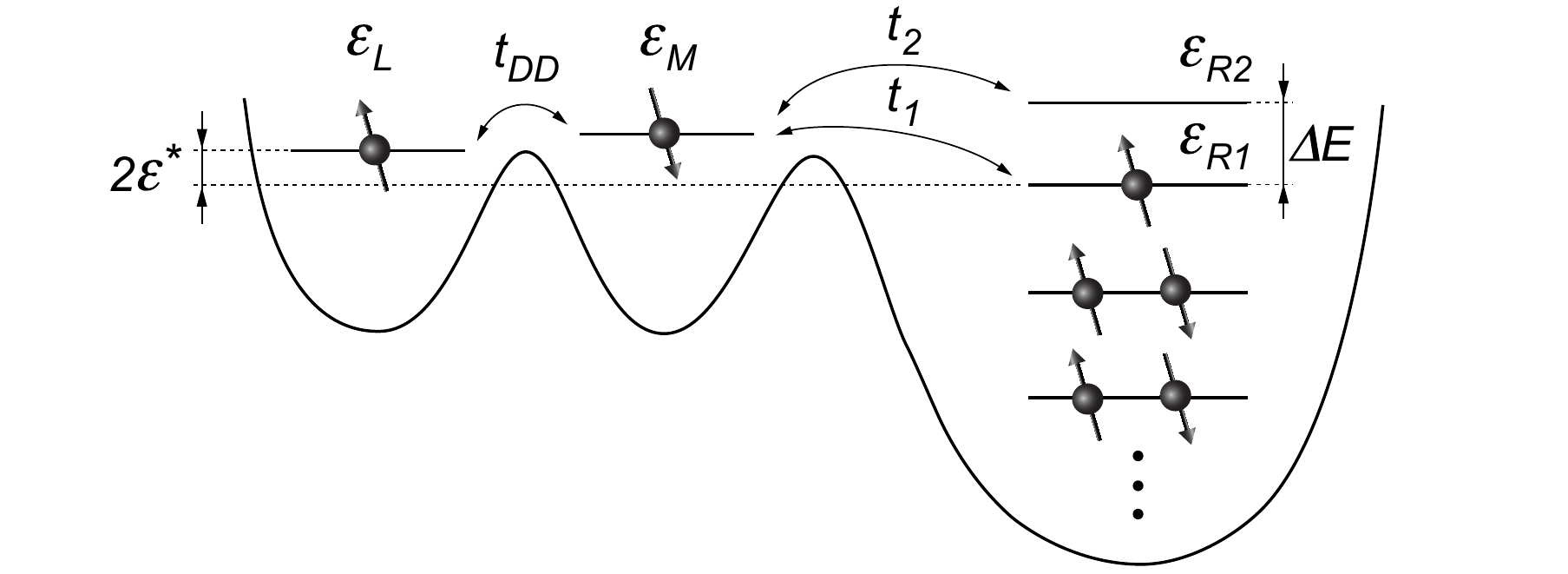}
	\caption{
	Schematic of a two-electron double quantum dot coupled to an odd-occupied spin-1/2 multielectron dot. Symbols $\varepsilon_{L/M/R1/R2}$ indicate single particle energies of the lowest orbitals in the double dot and the two lowest orbitals above the effective vacuum in the multielectron dot. Arrows indicate tunnel couplings between the left and middle orbital ($t_{DD}$) and between the middle orbital  and each of the two orbitals in the multielectron dot ($t_{1/2}$). The single-particle energy difference between the two orbitals on the multielectron dot is indicated by $\Delta E = \varepsilon_{R2} - \varepsilon_{R1}$. Detuning $\varepsilon^* = (\varepsilon_L-\varepsilon_{R1})/2$ is varied when calculating the energy diagram and leakage spectrum presented in Figs.~\ref{fig7}(b,e) and \ref{fig9}(b,e). 
	 Within the range of relevant parameters, low-laying orbitals in the right dot remain doubly occupied (i.e. spinless) and can be ignored when solving Eq. 
 (\ref{eq:spin12Ham}), thereby yielding a three-electron Hubbard model. [Similarly, a four-electron Hubbard model is solved to model the spin-1 case in Fig. ~\ref{fig13}(b,e).]}
	\label{fig6}
\end{figure}

\begin{figure}[tb]
	\centering
	\includegraphics[width=0.48\textwidth]{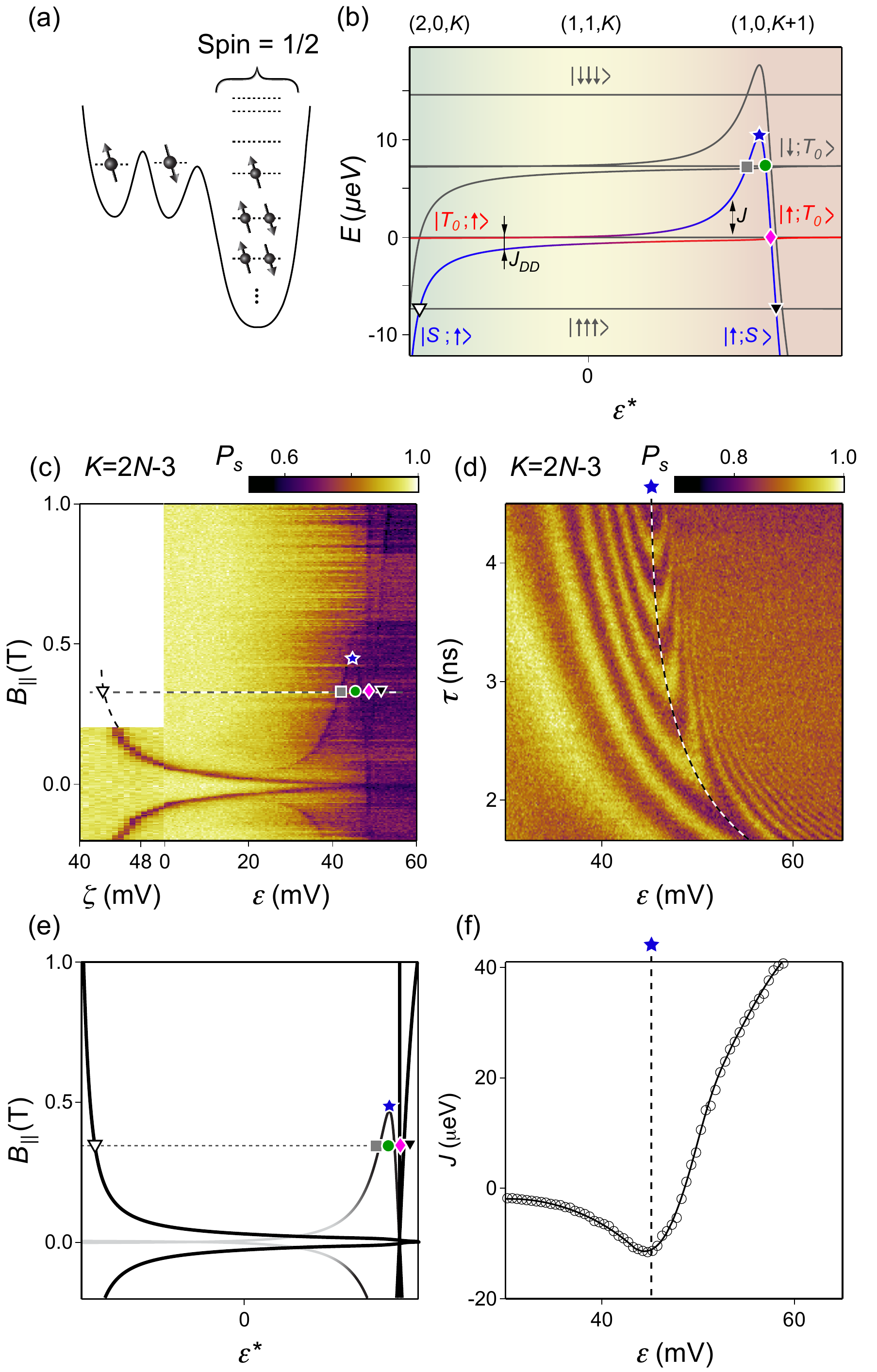}
	\caption{
	(a) Schematic of the odd-occupied multielectron quantum dot with spin-1/2 ground state, tunnel coupled to the two-electron double quantum dot.
	(b) The inferred energy diagram at the transition between $(2,0,2N \! - \! 3)$, $(1,1,2N \! - \! 3)$ and $(1,0,2N \! - \! 2)$ charge configurations, for a finite magnetic field $B_\parallel = 330$~mT. The markers indicate crossings revealed by the leakage spectroscopy measurement presented in panel (c).
	Energies are measured relative to the energy of the state $\ket{Q} \propto \ket{\uud}+\ket{\udu}+\ket{\duu}$.	
	(d) Time-resolved measurement of exchange oscillations between the $2N \! - \! 3$ occupied spin-1/2 multielectron dot and the middle electron.
	(e) Leakage spectrum expected from the Hubbard model.
	(f) Exchange profile $J(\varepsilon)$ extracted from the pattern of exchange oscillations in panel (d).
	}
	\label{fig7}
\end{figure}

\begin{figure}[tb]
	\centering
	\includegraphics[width=0.48\textwidth]{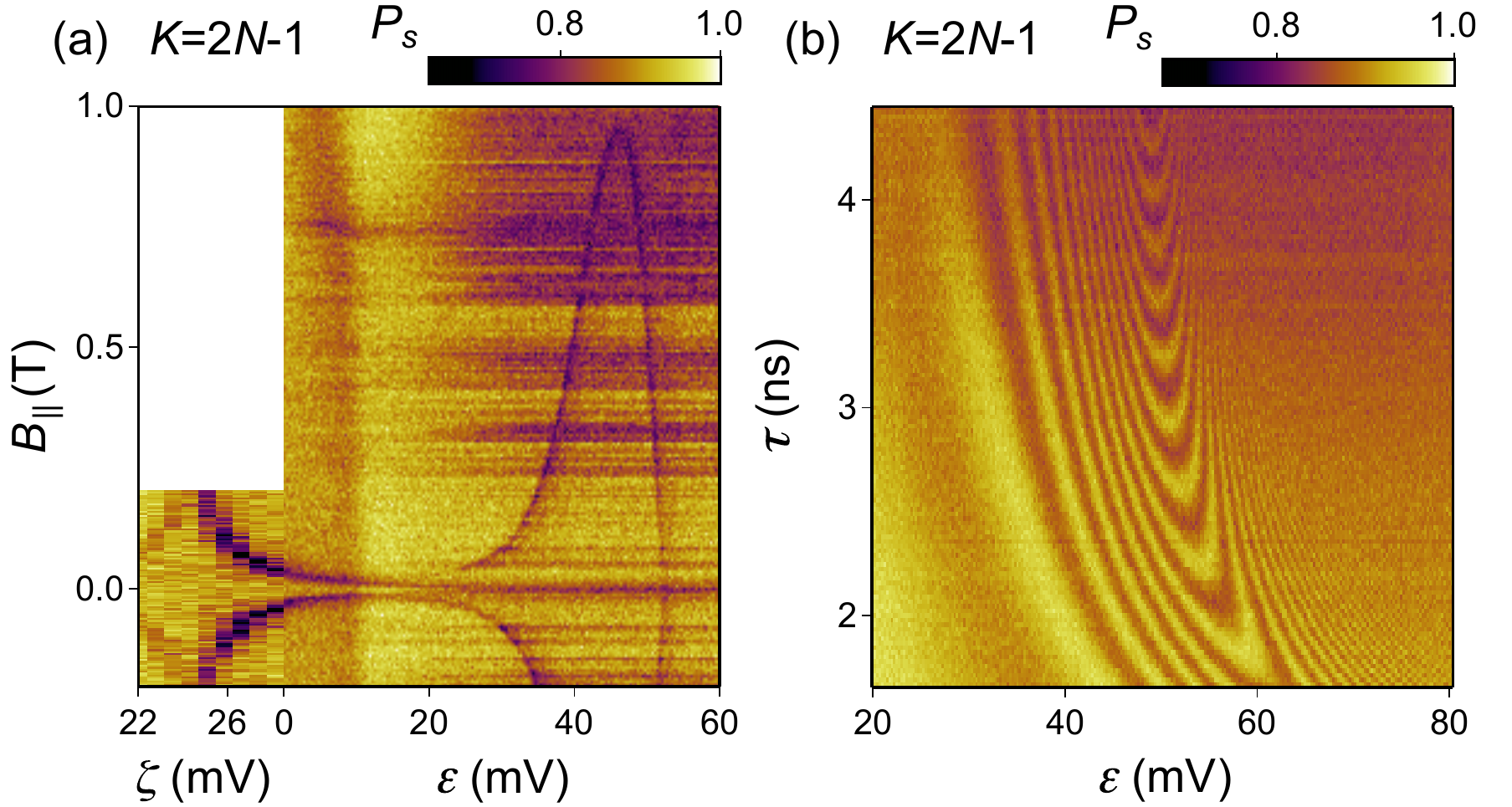}
	\caption{Leakage spectroscopy (a) and time-resolved exchange oscillations measurement (b) for the multielectron dot occupied by $2N \! - \! 1$ electrons.
	}
	\label{fig8}
\end{figure}

We now focus on two particular odd occupancies of the multielectron quantum dot, $2N \! - \! 3$ and $2N \! - \! 1$, that turn out to behave similar to each other but strikingly different than the (1,1,1) system considered in Subsection~\ref{subsec:TQD}. 
In these two cases the multielectron quantum dot has a single unpaired spin on the highest occupied orbital, while remaining electrons are paired up on lower laying orbitals as spin singlets [see schematic in Fig.~\ref{fig6}]. The data we describe below [Figs. \ref{fig7}(c,d) and \ref{fig8}] will be interpreted within the Hubbard model associated with Fig.~\ref{fig6}.

We first discuss leakage spectroscopy measurements for the multielectron quantum dot with $2N \! - \! 3$ occupancy. The left part of Fig.~\ref{fig7}(c) corresponds to a configuration in which the multielectron quantum dot is not significantly exchange coupled to the double quantum dot (i.e. $J \approx 0$). As expected from a conventional two-electron double dot, we observe in this regime a sharp feature of suppressed $P_S$ with a shape similar to the ``spin funnel'' presented in the left half of Fig. \ref{fig3}(c).
Assuming that the multielectron dot simply constitutes a spectator spin in this regime (which can be representated as $\ket{\uparrow}$ anticipating the following analysis), we can associate this feature with the crossing between states $\ket{S; \uparrow}$ and $\ket{\uparrow\uparrow;\uparrow}$ ($=\ket{\uuu}$). In this interpretation, the curvature of the ``spin funnel'' (marked by white triangle) reflects the gradual transition of the associated charge configuration from $(2,0,2N \! - \! 3)$ to $(1,1,2N \! - \! 3)$ occupancy.

For intermediate values of $\varepsilon$ the multielectron spin results in a leakage pattern that differs from the leakage spectrum of a conventional three-electron triple quantum dot (discussed in Subsection~\ref{subsec:TQD}). Namely, the line associated with the crossing between $\ket{S; \uparrow}$ and $\ket{\uuu}$ [white triangle in Fig.~\ref{fig7}(c)] converges towards $B_\parallel=0$. Meanwhile, a second sharp feature emerges. With increasing $\varepsilon$ it first shifts towards larger values of $B_\parallel$ (grey square), then reaches a maximum (blue star) before returning towards $B_\parallel=0$ (green circle). 
At the point where this feature crosses $B_\parallel=0$ we observe two additional sharp leakage features. The position of one of them is approximately independent on $B_\parallel$ (pink diamond) while the other feature diverges towards large $B_\parallel$ for increasing values of $\varepsilon$ (black triangle). This non-trivial leakage spectrum occurs at a detuning ($\varepsilon\gtrsim30$~mV) where the charge state of the ground state transitions into $(1,0,2N \! - \! 2)$. 

To explain this peculiar leakage pattern, we modify the Hubbard model of Eq.~\ref{eq:spin0Ham} that successfully described the even-occupancy spin-0 case.  Specifically, we now assume that the gate voltage of the multielectron quantum dot has been tuned such that the ground state is odd-occupied, and is effectively described by a single unpaired spin on orbital $R1$ (illustrated in Fig.~\ref{fig6}). Lower lying orbitals are assumed to be occupied by spinless electron pairs and are ignored. However, we found it necessary to include a higher laying empty orbital $R2$, $\Delta E$ higher compared to $R1$. 
Including the spin correlation term $\xi$ of Eq.~\ref{eq:JB}, this generalizes Eq.~\ref{eq:spin0Ham} to the following Hamiltonian, appropriate for the system illustrated in Fig.~\ref{fig6}:

\begin{align}
	\hat{H}_{\text{spin-1/2}} &= \sum_{i=L,M,R} \left( \varepsilon_i \hat{n}_i + \frac{U_i}{2}\hat{n}_i(\hat{n}_i-1) \right) + \sum_{i\neq j}\frac{K_{ij}}{2}\hat{n}_i\hat{n}_j \nonumber \\
	&- \frac{\xi}{2} \hat{S}^2 \nonumber \\
    &- t_{DD} \sum\limits_{\alpha=\uparrow,\downarrow}  (\hat{c}_{L,\alpha}^\dag \hat{c}_{M,\alpha} + \hat{c}_{M,\alpha}^\dag \hat{c}_{L,\alpha}) \nonumber \\
	&- t_1\sum\limits_{\alpha=\uparrow,\downarrow}  (\hat{c}_{M,\alpha}^\dag \hat{c}_{R1,\alpha} + \hat{c}_{R1,\alpha}^\dag \hat{c}_{M,\alpha}) \nonumber \\
    &- t_2\sum\limits_{\alpha=\uparrow,\downarrow}  (\hat{c}_{M,\alpha}^\dag \hat{c}_{R2,\alpha} + \hat{c}_{R2,\alpha}^\dag \hat{c}_{M,\alpha})\, .
    \label{eq:spin12Ham}
\end{align}
The first line of this equation captures the gate-tunable chemical potentials and Coulomb interactions, and hence it is diagonal in terms of the spin occupancy numbers. 
The second line, proportional to $\xi$, captures the spin correlation energy: it is a phenomenological term that favors a $S=1$ triplet configuration when both levels $R1$ and $R2$ are occupied. Here the operator of the multielectron quantum dot spin in orientation $j=x,y,z$ is
$\hat{S}^j = \tfrac{1}{2} \sum_{\lambda,\alpha,\alpha'} \hat{c}_{\lambda,\alpha}^\dag \sigma^j_{\alpha,\alpha'} \hat{c}_{\lambda,\alpha'}$, where $\lambda=R1, R2$.
This term is important when the single particle spacing  $\Delta E \equiv \varepsilon_{R2}-\varepsilon_{R1}$ is relatively small. 
The remaining terms proportional to $t_{DD}$, $t_1$, and $t_2$ are chosen to be real and positive. They describe, respectively, tunnel couplings within the double quantum dot, between the middle dot $M$ and $R1$, and between $M$ and $R2$.

In Fig.~\ref{fig7}(b) we present the energy diagram of the double quantum dot coupled to the spin-1/2 multielectron quantum dot, calculated from the 3-electron spectrum of Eq.~\ref{eq:spin12Ham} using $t_1=12$~$\mu$eV, $t_2=48$~$\mu$eV, $t_{DD}=12$~$\mu$eV and $\Delta E = 160$~$\mu$eV. All plotted energies are measured relative to the energy of the state $\ket{Q} \propto \ket{\uud}+\ket{\udu}+\ket{\duu}$, which appears  at $E=0$. In such a plot, triplet-like states display constant energies, whereas singlet-like states depend on detuning.

Again we have inspected two relevant states (marked blue for a singlet-like state and red for a triplet-like state) in more detail, and indicate their spin states in the limit of very negative and very positive detunings. Further, for negative $\varepsilon^*$, we have labeled their energy splitting by $J_{DD}$, to indicate that this exchange coupling arises predominantly from interdot tunneling within the double dot (discussed in next paragraph). For positive $\varepsilon^*$, interdot tunneling within the double dot is negligible, but tunneling between the middle dot and the multielectron dot is important. The resulting exchange coupling, labeled $J$, reflects non-trivial spin-correlation effects arising from the orbitals within the multielectron dot (discussed below). 

The left part of Fig.~\ref{fig7}(b), characterized by finite $J_{DD}$ and negligible $J$, has an interpretation very similar to the left part of Fig.~\ref{fig5}(b), i.e., a three-electron triple dot with finite $J_L$ and negligible $J_R$. In particular, the crossing between $\ket{S; \uparrow}$ and $\ket{\uuu}$ states (white triangle) is expected to result in a ``spin funnel''-like feature in this regime, for the same reasons as in Subsection~\ref{subsec:TQD}. 

For increasing values of $\varepsilon^*$, however, the singlet-like state $\ket{S; \uparrow}$ continuously changes its spin texture from $\ket{\uparrow\downarrow\uparrow} - \ket{\downarrow\uparrow\uparrow}$ to $\ket{\uparrow\downarrow\uparrow} + \ket{\uparrow\uparrow\downarrow}$, becoming a triplet-like state $ \ket{\uparrow;T_0}$ (we omit normalization). 
Concurrently, the triplet-like state $\ket{T_0;\uparrow}$ continuously changes its spin texture from $\ket{\uparrow\downarrow\uparrow} + \ket{\downarrow\uparrow\uparrow}$ to $\ket{\uparrow\downarrow\uparrow} - \ket{\uparrow\uparrow\downarrow}$, becoming a singlet-like state $\ket{\uparrow;S}$ for large $\varepsilon^*$.
In the model, this transition is driven by the \emph{negative} exchange interaction arising from $\xi=0.1$ meV in conjunction with large tunneling $t_2$ to the second orbital [cf. Eq. (\ref{eq:spin12Ham})], which increases the energy of the singlet-like state $\ket{\uparrow;S}$ relative to the triplet-like state $\ket{\uparrow;T_0}$.
However, for even larger $\varepsilon^*$ [i.e. in the $(1,0,K \! + \! 1)$ charge configuration] the tunneling effects become suppressed, and hence the singlet-like state $\ket{\uparrow;S}$ becomes the ground state due to a relatively large level spacing $\Delta E > \xi$. 

The negative sign of the exchange interaction $J$ for intermediate values of $\varepsilon^*$ explains why for Zeeman splittings smaller than the maximum energy (blue star) two additional crossings are expected (marked in Fig.~\ref{fig7}(b) by gray square and green circle), consistent with features observed in the leakage spectrum of Fig.~\ref{fig7}(c). 
For both crossings we expect leakage from $\ket{\uparrow; S}$ into $\ket{\uparrow;T_-}$, as this state has total spin projection $S_z=-1/2$ and is accessible via electron-nuclear flip-flop processes. 

In the context of these results from the Hubbard model [Fig.~\ref{fig7}(b)], we are able to return to the measurements [Fig.~\ref{fig7}(c,d) and Fig.~\ref{fig8}]  and discuss a few more details. 

For sufficiently large values of $\varepsilon^*$ in Fig.~\ref{fig7}(b), the energy of the singlet-like state $\ket{\uparrow; S}$ decreases and becomes lower than the energy of the triplet-like state $\ket{\uparrow;T_0}$. 
For the crossing of these two states we expect a leakage feature (indicated by the pink diamond), at a detuning value that is independent of the magnetic field (since the two involved states have the same spin projection).
For higher detuning the energy of the $\ket{\uparrow;S}$ state further decreases, and crosses the $\ket{\uuu}$ state, resulting in the leakage feature indicated by a black triangle. 
Indeed, both leakage features are clearly observed in the experiment, as indicated by the pink diamond and black triangle in Fig.~\ref{fig7}(c). 
In particular, the divergence of one leakage feature for increasing $\varepsilon$ (black triangle) implies that the multielectron quantum dot in $2N \! - \! 2$ occupancy has a spin-0 ground state, consistent with the evidence presented in Sec.~\ref{sec:0}. 

The measured leakage spectrum does not reveal the crossing between $\ket{\uparrow;S}$ and the fully polarized $\ket{\ddd}$ state. This is expected, as leakage into the $\ket{\ddd}$ state ($S_z=-3/2$) would require a change of the electronic spin projection by 2 (which is not expected for weak spin-orbit interaction and typical Overhauser field gradients).

An identical analysis of the leakage spectroscopy measurements for the $2N \! - \! 1$ occupancy of the multielectron quantum dot [Fig.~\ref{fig8}(a)] yields the same conclusion. In particular, it indicates that the multielectron dot with $2N$ occupancy has a spin-0 ground state, in agreement with the evidence presented in Sec.~\ref{sec:0}.

Similar to our procedure in Sec.~\ref{sec:0}, we extract the relevant state crossings from calculations as in Fig.~\ref{fig7}(b) for varying external magnetic fields, and thereby generate the leakage spectrum expected for the system of Fig.~\ref{fig6}.  The resulting leakage map, shown in Fig.~\ref{fig7}(e), qualitatively reproduces all features of the leakage spectroscopy measurements. 
This calculation also predicts leakage between the triplet-like state $\ket{T_0;\uparrow}$ and a state with $S=+3/2, S_z=+3/2$, which we plot in light gray color.
This leakage feature is not apparent in the measured data, as our initialization pulses were designed to prepare the $\ket{S;\uparrow}$ state, which is orthogonal to the states that anticross.

The interpretation of the maximum in Fig.~\ref{fig7}(c) (blue star) as an extremum in $J(\varepsilon)$ can be confirmed directly in the time domain, by reducing the interaction time $\tau$ and inspecting coherent oscillations between the singlet-like and triplet-like state at intermediate values of $\varepsilon$. In this technique, the oscillation frequency observed at detuning $\varepsilon$ is a quantitative measure for $|J(\varepsilon)|$~\cite{Laird2010}. The experimental data for $2N \! - \! 3$ and $2N \! - \! 1$ occupancy of the multielectron dot are presented in Figs.~\ref{fig7}(d) and \ref{fig8}(b), respectively. In both cases, an increase followed by a decrease in oscillation frequency with increasing detuning is clearly observed. 
For comparison, the exchange energy extracted from the leakage spectroscopy pattern for $2N \! - \! 3$ occupancy is presented in Fig.~\ref{fig7}(f).
 We observe that the minimum of $J(\varepsilon)$ in Fig.~\ref{fig7}(f) occurs at that value of $\varepsilon$ for which the oscillation frequency in Fig.~\ref{fig7}(d) shows a maximum (for large $\tau$ this agreement is good, whereas for small $\tau$ pulse distortions arising from finite-rise-time effects associated with our cryostat wiring become signifiant). The overall agreement between maxima in leakage spectra [blue star in Figs.~\ref{fig7}(c) and \ref{fig8}(a)] and maxima in oscillation speed [Figs.~\ref{fig7}(d) and \ref{fig8}(b)] confirms that the exchange interaction strength has an extremum as a function of $\varepsilon$. 
Although the oscillations in Figs.~\ref{fig7}(d) and \ref{fig8}(b) do not reveal the absolute sign of the exchange coupling, these measurements do confirm its \emph{change} of sign.

To summarize, the qualitative agreement between observed and expected features leads us to accept the  physical inspection of the Hubbard model results [Fig.~\ref{fig7}(b)] as the correct interpretation of the measurement results [Fig.~\ref{fig7}(c,d) and Fig.~\ref{fig8}]. 
This allows us to conclude that for intermediate values of $\varepsilon$ the triplet-like configuration associated with $\ket{\uparrow;T_0}$ has a \emph{lower} energy than the singlet-like state associated with $\ket{\uparrow;S}$. We refer to this inversion as \emph{negative} exchange coupling.
Both for the $2N \! - \! 3$ and $2N \! - \! 1$ occupancy of the multielectron quantum dot we observe this negative (i.e., triplet-preferring) exchange coupling to the proximal  electron spin, as long as the proximal spin resides in the middle dot, i.e. for charge configurations $(1,1,2N \! - \! 3)$ or  $(1,1,2N \! - \! 1)$.  Once a sufficiently large detuning voltage transfers the proximal electron onto the multielectron dot, i.e. resulting in a $(1,0,2N \! - \! 2)$ or $(1,0,2N)$ charge configuration, the exchange interaction becomes positive, consistent with a spin-0 ground state for the even-occupied multielectron quantum dot (studied in Sec.~\ref{sec:0}).

\subsection{Negative exchange within the multielectron dot  ($K=2N \! + \! 1$)}
\label{subsec:negative-J}

\begin{figure}[tb]
	\centering
	\includegraphics[width=0.48\textwidth]{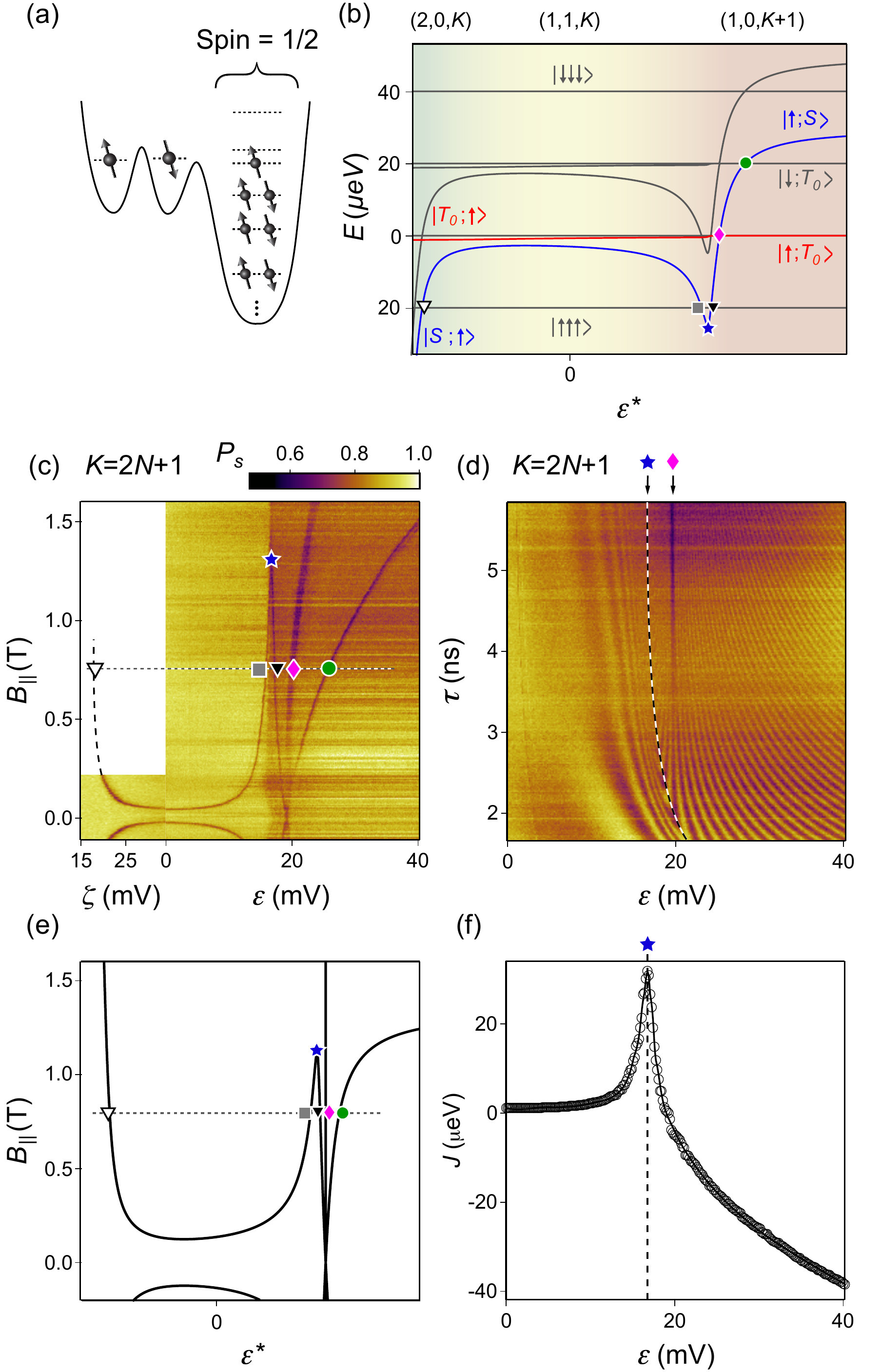}
	\caption{
	(a) Schematic of the odd-occupied multielectron quantum dot with spin-1/2 ground state, tunnel coupled to the two-electron double quantum dot.
	(b) Calculated energy diagram at the transition between $(2,0,2N \! + \! 1)$, $(1,1,2N \! + \! 1)$ and $(1,0,2N \! + \! 2)$ charge configurations, for a finite magnetic field $B_\parallel = 900$~mT and input parameters motivated by the observed spectrum in panel (c). Markers indicate crossings revealed by the leakage spectroscopy measurement presented in panel (c).
	Energy are measured relative to the energy of the state $\ket{Q} \propto \ket{\uud}+\ket{\udu}+\ket{\duu}$.	
	(d) Time-resolved measurement of exchange oscillations between the $2N \! + \! 1$-occupied spin-1/2 multielectron quantum dot and the middle electron. Colorscale as in (c).
	(e) Leakage spectrum expected for the Hubbard model.
	(f) Exchange profile $J(\varepsilon)$ extracted from the pattern of exchange oscillations in panel (d).
	}
	\label{fig9}
\end{figure}

Next we focus on the $2N \! + \! 1$ occupation of the multielectron dot. Similarly to the odd occupancies $2N \! - \! 3$ and  $2N \! - \! 1$ (Subsection~\ref{subsec:positive-J}) we expect that a single unpaired electron occupies the highest occupied orbital of the multielectron dot [Fig.~\ref{fig9}(a)]. However, the leakage spectroscopy measurement (Fig.~\ref{fig9}(c), discussed below) implies that the exchange interaction with the neighboring middle electron is qualitatively different. By changing the model parameters associated with Fig.~\ref{fig6} slightly (in particular the relative magnitudes of $t_1$, $t_2$, $\Delta E$ and $\xi$) the calculated spectrum can be made to match the experimental data.  

The left-hand side of Fig.~\ref{fig9}(c) presents an experimental leakage spectrum similar to that observed for $2N \! - \! 3$ and  $2N \! - \! 1$, and hence we tentatively associate the sharp funnel-like leakage feature (white triangle) with the crossing between singlet-like state $\ket{S; \uparrow}$ and the fully polarized state $\ket{\uuu}$. 
In contrast to the previous cases, this feature does not converge to $B=0$ for increasing $\varepsilon$, but increases to high magnetic fields (grey square) for intermediate values of $\varepsilon$. This increase occurs within the $(1,1,2N \! + \! 1)$ charge state, indicating that the singlet-like state $\ket{\uparrow; S}$ has a lower energy than the triplet-like state as long as the proximal electron resides on the middle dot. For larger detuning, around a charge transition to $(1,0,2N \! + \! 2)$ charge state, this leakage feature reaches a (sharp) maximum and then crosses through $B_\parallel=0$ (pink diamond), along with two sharp leakage features appearing.

This leakage spectrum can be reproduced within our model (Fig.~\ref{fig6}) using $t_1=30$~$\mu$eV, $t_2=6$~$\mu$eV,  $t_{DD}=20$~$\mu$eV and $\Delta E = 30$~$\mu$eV. 
The calculated energy diagram, obtained from the Hamiltonian~\eqref{eq:spin12Ham}, is presented in Fig.~\ref{fig9}(b). For these parameters, the spectrum reveals a positive (singlet-preferring) exchange interaction for the $(1,1,2N \! + \! 1)$ configuration and a negative (triplet-preferring) exchange interaction for $(1,0,2N \! + \! 2)$. Moreover, we can associate all leakage features observed in Fig.~\ref{fig9}(c) with specific crossings in the calculated spectrum. In particular, the three sharp leakage features converging towards $B_\parallel=0$ correspond to the crossings between $\ket{\uparrow;S}$ and states with total spin $S=3/2$ and spin projection, respectively, $S_z=+3/2$ (black triangle), $+1/2$ (pink diamond) and $-1/2$ (green circle).

Following the reasoning from Subsection~\ref{subsec:positive-J} we come to the conclusion that the $2N \! + \! 2$ occupied multielectron dot has a spin-1 ground state. Indeed, in Sec.~\ref{sec:1} we will present exchange effects in the $(1,1,2N \! + \! 2)$ system that are consistent with a spin-1 ground state of the multielectron dot.

In Fig.~\ref{fig9}(d) we present time-resolved exchange oscillations measured for the same configuration as for the leakage spectroscopy in Fig.~\ref{fig9}(c). As for the $2N \! - \! 3$ and $2N \! - \! 1$ occupancies we find that the oscillation frequency reaches a maximum for the same value of $\varepsilon$ as the local maximum in the leakage spectrum (blue star). [Deviations appear for short values of $\tau$, due to finite-rise-time effects as in Fig.~\ref{fig7}(d) and \ref{fig8}(b)].  For the highest values of $\tau$ we additionally observe a suppression of $P_S$ at $\varepsilon \approx 21$~mV, which we attribute to the onset of incoherent leakage from the singlet-like state into the $\ket{\uparrow; S}$ state (pink diamond). 

More details about the charge occupancy presented in this subsection ($K=2N \! + \! 1$), including tunability of the exchange profile, can be found in Ref.~\onlinecite{Negative-J}.

\subsection{Other odd occupancies of the multielectron quantum dot ($K=2N \! - \! 5$, $K=2N \! + \! 3$)}

\begin{figure}[tb]
	\centering
	\includegraphics[width=0.48\textwidth]{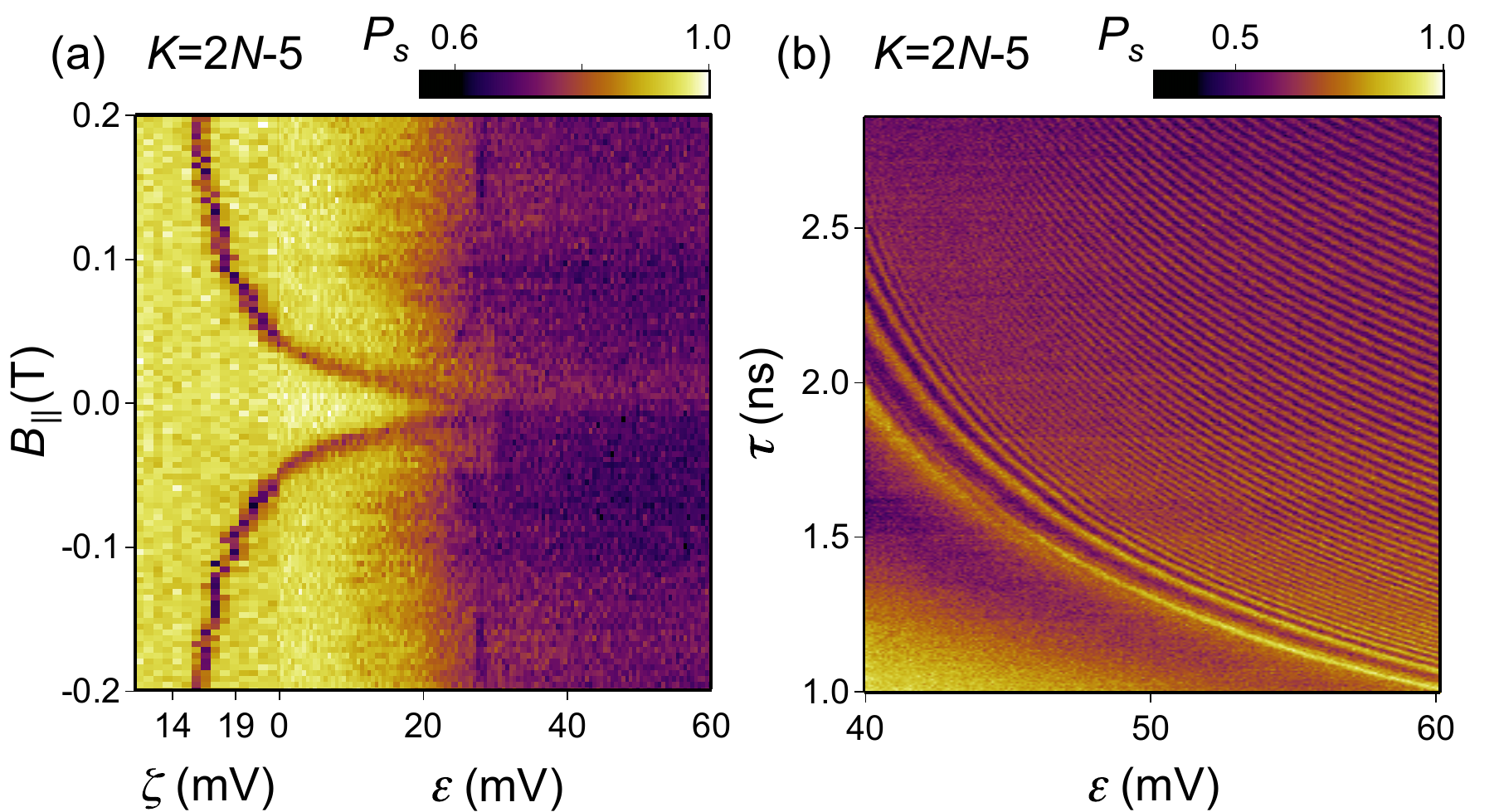}
	\caption{Leakage spectroscopy (a) and time-resolved exchange oscillations (b) for the multielectron dot occupied by $2N \! - \! 5$ electrons.
	}
	\label{fig10}
	\vspace{10pt}
%
	\includegraphics[width=0.48\textwidth]{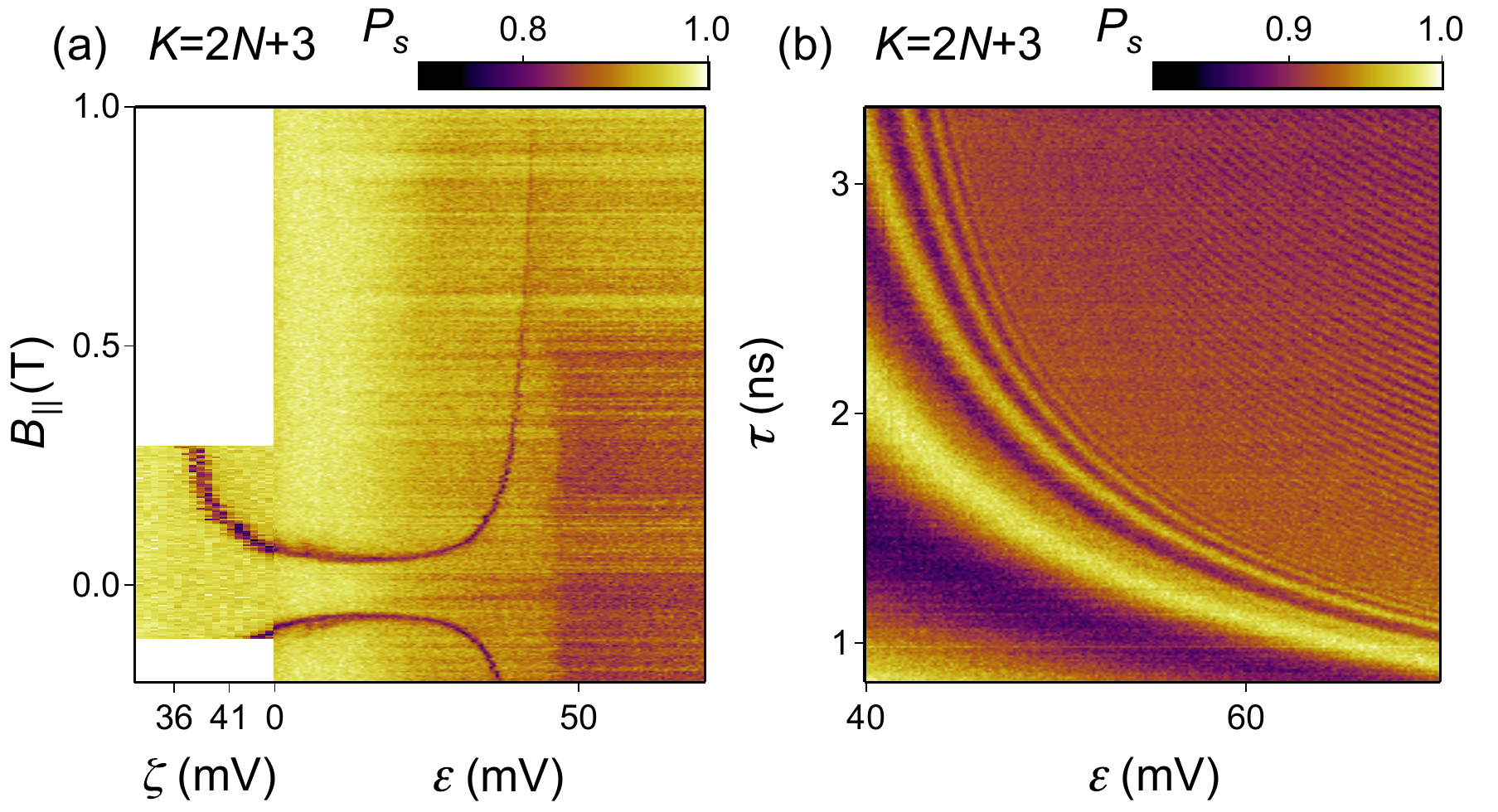}
	\caption{Leakage spectroscopy (a) and time-resolved exchange oscillations (b) for the multielectron dot occupied by $2N \! + \! 3$ electrons.
	}
	\label{fig11}
\end{figure}

We now present the results of the leakage spectroscopy and exchange oscillation measurements for the $2N \! - \! 5$ and $2N \! + \! 3$ occupancies of the multielectron dot. These are the most extreme occupancies studied in this work. Further addition or removal of electrons is possible, but would require significant changes of several tuning voltages to maintain a useful tunnel coupling between the middle and multielectron dot, presumably due to considerable changes of the quantum dot area.
The observed behavior is consistent with the ground states found for the other odd occupations, namely a spin-1/2 multielectron dot, but the reduced quality of data did not allow an analysis as detailed as that in Subsections~\ref{subsec:positive-J} and \ref{subsec:negative-J}.

In Fig.~\ref{fig10}(a) we present leakage spectroscopy for the $2N \! - \! 5$ occupancy of the multielectron quantum dot. We attribute the two funnel-like features in the left half of the panel with the usual exchange coupling within the two-electron double quantum dot, decreasing in strength with increasing $\varepsilon$. However, at $\varepsilon \approx 20$~mV each feature appears to split into two features. One converges towards $B=0$ while the other quickly increases and possibly reaches a maximum at $\varepsilon \approx 28$~mV before returning and crossing $B=0$ at about $\varepsilon = 30$~mV. 
On the one hand, this may indicate that the exchange interaction strength between the middle electron and the spin-1/2 multielectron dot has a negative sign for small wavefunction overlap [i.e. in the $(1,1,2N \! - \! 5)$ charge configuration] and positive for large wavefunction overlap [i.e. in the $(1,0,2N \! - \! 4)$ charge configuration]. On the other hand, the opposite behavior (i.e. exchange sign going from positive to negative) would also be consistent with the leakage pattern, although we dismiss this possibility based on the spin-0 behavior presented for $K=2N \! - \! 4$.

Exchange oscillations for $2N \! - \! 5$ occupancy are presented in Fig.~\ref{fig10}(b). Notably, there is no indication for a local maximum in the oscillation frequency, in contrast to an extremum in the exchange interaction strength that we inferred from the leakage spectrum in Fig.~\ref{fig10}(a). We do not understand the absence of an extremum in Fig.~\ref{fig10}(b), but note that the presence of exchange oscillations by itself is evidence for a spinful ground state of the $2N \! - \! 5$ occupied multielectron dot.

Leakage spectroscopy performed for the $2N \! + \! 3$ occupation, presented in Fig.~\ref{fig11}(a), reveals characteristics similar to those of the conventional three-electron triple quantum dot (Subsection~\ref{subsec:TQD}). This similarity, and the absence of unusual leakage features at the $(1,1,2N \! + \! 3)$ to $(1,0,2N \! + \! 4)$ charge transition, suggests that the multielectron quantum dot with this occupancy behaves as an ordinary spin-1/2 dot. However, we cannot fully exclude the possibility that at high $\varepsilon$ the exchange interaction reaches a maximum and possibly changes sign, as such a behavior is hard to detect for large tunnel couplings~\cite{Negative-J}.  In addition, the observed pattern of exchange oscillations in Fig.~\ref{fig11}(b) is not quite clear enough to support the presence or absence of a extremum in the exchange profile, due to an increased dephasing rate at the interdot charge transition~\cite{Dial2013}.

\subsection{Different exchange profiles for a spin-1/2 multielectron dot expected from the Hubbard model}
\label{subsec:phase_diagram}

\begin{figure}[tb]
	\centering
	\includegraphics[width=0.5\textwidth]{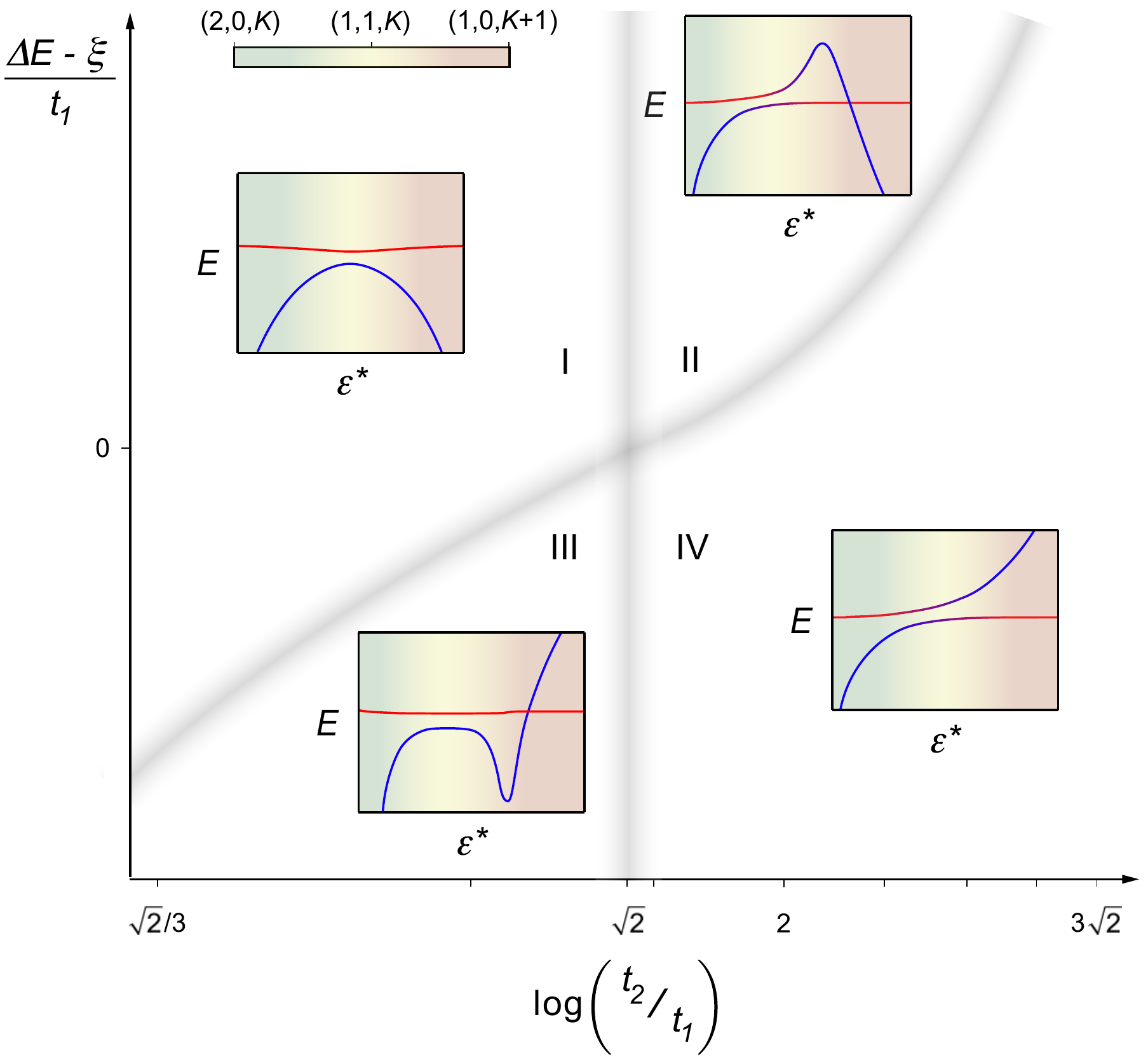}
	\caption{
	Illustration of qualitatively different exchange profiles arising from the interplay between the level spacing in the multielectron quantum dot $\Delta E$, spin correlation energy $\xi$ and tunnel couplings between a single-electron dot and two lowest orbitals of the multielectron quantum dot $t_{1/2}$. Colored lines in the insets I-IV represent the energies of the three-spin states with $S=1/2$, $S_z=-1/2$ as a function of detuning $\varepsilon^*$.
	}
	\label{fig12}
\end{figure}

According to our phenomenological model the effective exchange coupling between a spin-1/2 ground state of the multielectron dot and the middle spin depends on the precise choice of the various input parameters. However, the general behavior of the exchange profile falls into four main regimes as shown schematically in Fig.~\ref{fig12}.

In regime I the effective exchange coupling is always positive (singlet preferring) as $\varepsilon^*$ is tuned towards the charge transition where an additional electron moves onto the multielectron dot. The behavior is qualitatively similar to that of a three-electron triple dot ($K=1$, Subsection \ref{subsec:TQD}). 

In regime II, the effective exchange coupling is negative (triplet preferring) in the vicinity of the charge transition from $(1,1,K)$ to $(1,0, K \! + \! 1)$, but becomes positive again for higher detuning (i.e. the exchange profile includes a maximum and a zero crossing). This is the regime observed for $2N \! - \! 3$ and $2N \! - \! 1$ occupancies (Subsection \ref{subsec:positive-J}). 

Regime III is similar to Regime I for low detuning, including positive exchange at the charge transition from $(1,1,K)$ to $(1,0,K \! + \! 1)$, but shows a sign reversal followed by negative exchange for higher detuning, as observed for $2N \! + \! 1$ occupancy (Subsection \ref{subsec:negative-J}).

Regime IV is characterized by a negative exchange coupling that already develops at the charge transition from $(1,1,K)$ to $(1,0,K \! + \! 1)$, and the exchange coupling remains negative for higher detuning. Note that there is no zero crossing in the exchange profile, even though the exchange coupling for low detuning (corresponding to exchange between middle dot and left dot) is positive, and exchange and high detuning (corresponding to exchange between middle dot and the multielectron dot) is negative. This is because in the charge configuration $(1,1,K)$, tunneling across left and right barrier are both present, and hence eigenstates cannot be decomposed into a product state of one spectator spin (in either left or right dot) and a remaining spin-singlet (or spin triplet) state. In other words, the energy spitting between eigenstates is non-zero at this detuning, but cannot be classified as positive or negative because the eigenstates themselves are superpositions of singlet-like (blue) and triplet-like (red) states. 

In Figure~\ref{fig12} we intentionally omit units on the four insets, to emphasize that our simple theoretical model predicts four qualitatively different regimes of exchange profiles. However, only regimes II and III have been observed in our multielectron device (regime I has been observed only for $K=1$). To gain insight into the physics that -- within the theoretical model -- gives rise to these four regimes, we can inspect the crossover between these regimes in more detail. 
In particular, we can identify the boundaries between these four regimes by analyzing the role of two dimensionless quantities.  

The first parameter is $(\Delta E - \xi)/t_1$, and constitutes the vertical axis of Fig.~\ref{fig12}. This parameter can be positive or negative, depending on the relative strength of the spin correlations.  When positive, i.e., $\Delta E > \xi$, the energy separation of the two relevant single-particle levels 1 and 2 in the multielectron dot is larger than the spin correlation energy, thereby suppressing the formation of a high-spin ground state. Accordingly, if the middle electron is transferred into the right dot (large detuning), its lowest energy state will be a singlet configuration with level 1 doubly occupied. When this parameter is negative, i.e., $\Delta E < \xi$, the spin correlation energy is larger than the kinetic energy required to form a high-spin state, and hence a triplet configuration with an electron in both level 1 and level 2 is energetically preferred past the charge transition.  

Near the charge transition the effective exchange coupling between the multielectron dot and the double quantum dot results from a competition between a positive (singlet-preferring) contribution and a negative (triplet-preferring) contribution. These contributions arise from virtual transitions to the two doubly occupied configurations of the multielectron dot. The relative size of these contributions depends on the tunnel couplings $t_1$ and $t_2$, and consequently the second parameter we use to describe the spectrum is the ratio $t_2/t_1$. This parameter forms the horizontal axis of Fig.~\ref{fig12}.

For small detuning, i.e. when the charge configuration of the ground state is predominantly $(1,1,K)$, the Hubbard model can be analyzed perturbatively. In this regime the positive contribution to the exchange is approximately $J_S=2 t_1^2/\Delta E_S$ where $\Delta E_S$ is the difference in energy to the doubly occupied singlet state. $\Delta E_S$ changes with $\varepsilon^*$ and decreases as the system is tuned towards the charge transition. The negative contribution is approximately $J_T=- t_2^2/\Delta E_T$ where $\Delta E_T$ is the difference in energy to the doubly occupied triplet state. (Note that $\Delta E_T-\Delta E_S= \Delta E - \xi$.) Far from the charge transition (i.e. deep inside the $(1,1,K)$ configuration) we have $\Delta E_S\simeq \Delta E_T$, and hence we expect negative exchange for $|J_T|\gtrsim J_S$ if $t_2\gtrsim \sqrt{2}t_1$. Accordingly, regimes II and IV, which are characterized by the negative exchange as the charge transitions from $(1,1,K)$ to $(1,0,K \! + \! 1)$, occur in the region $t_2/t_1\gtrsim \sqrt{2}$ (Fig.~\ref{fig12}).

Now lets consider the region where $(\Delta E - \xi)/t_1>0$ so that $\Delta E_S< \Delta E_T$. When $t_2/t_1\lesssim \sqrt{2}$ we expect that $J_S\gtrsim |J_T|$ for all $\varepsilon^*$. This means that the effective exchange is always positive,  and indeed Fig.~\ref{fig12} shows behavior I in this range of parameters. On the other hand, when $t_2/t_1\gtrsim \sqrt{2}$ we have $|J_T|\gtrsim J_S$ far from the charge transition. As one approaches the charge transition one might naively expect that $\Delta E_S$ eventually becomes very small (specifically $\Delta E_S \ll \Delta E_T$), which would imply $J_S > |J_T|$. In other words, the effective exchange would change sign, placing the system in regime II. However, we find that when $(\Delta E - \xi)/t_1$ becomes sufficiently small the system actually transitions into regime IV (i.e. the effective exchange remains negative all the way to the charge transition). This demonstrates that perturbative expressions for the exchange splitting should not be trusted close to the charge transition. 
Specifically, near $\Delta E_S\simeq t_1$ significant corrections are needed, and from the full Hubbard model we expect that $J_S$ will not keep increasing beyond this point (in contradiction with the perturbative intuition). 
Therefore, a more accurate location for the crossover between regime II and regime IV is the locus of points where $J_S=|J_T|$ when $\Delta E_S \simeq t_1$. This boundary is indicated in~Fig.~\ref{fig12}, and we verified that spectra of the full Hubbard model are in agreement with this choice.

Finally, the boundary between regime I and regime III sits in the region where $(\Delta E - \xi)/t_1<0$ and so $\Delta E_T< \Delta E_S$. Following the same logic as for the boundary between II and IV, the crossover between regime I and regime III lies close to the locus of points where $J_S=|J_T|$ when $\Delta E_T \simeq t_2$.

Curiously, all exchange profiles observed for the multielectron quantum dot fall into regime II or III. Possibly this is related to an electron occupying an higher energy orbital (i.e. larger kinetic energy) having an increased tendency to penetrate the potential barrier between the dots (i.e. resulting in $t_2>t_1$). However, the number of occupancies studied in this work is too small to draw any general conclusions. To resolve this question, it would be beneficial to investigate more devices or to use distorting gates\cite{Folk1996} to change the quantum dot potential and thereby gather meaningful statistics.

\section{Spin-1 behavior for $K=2N \! + \! 2$}
\label{sec:1}

\begin{figure}
	\includegraphics[width=0.5\textwidth]{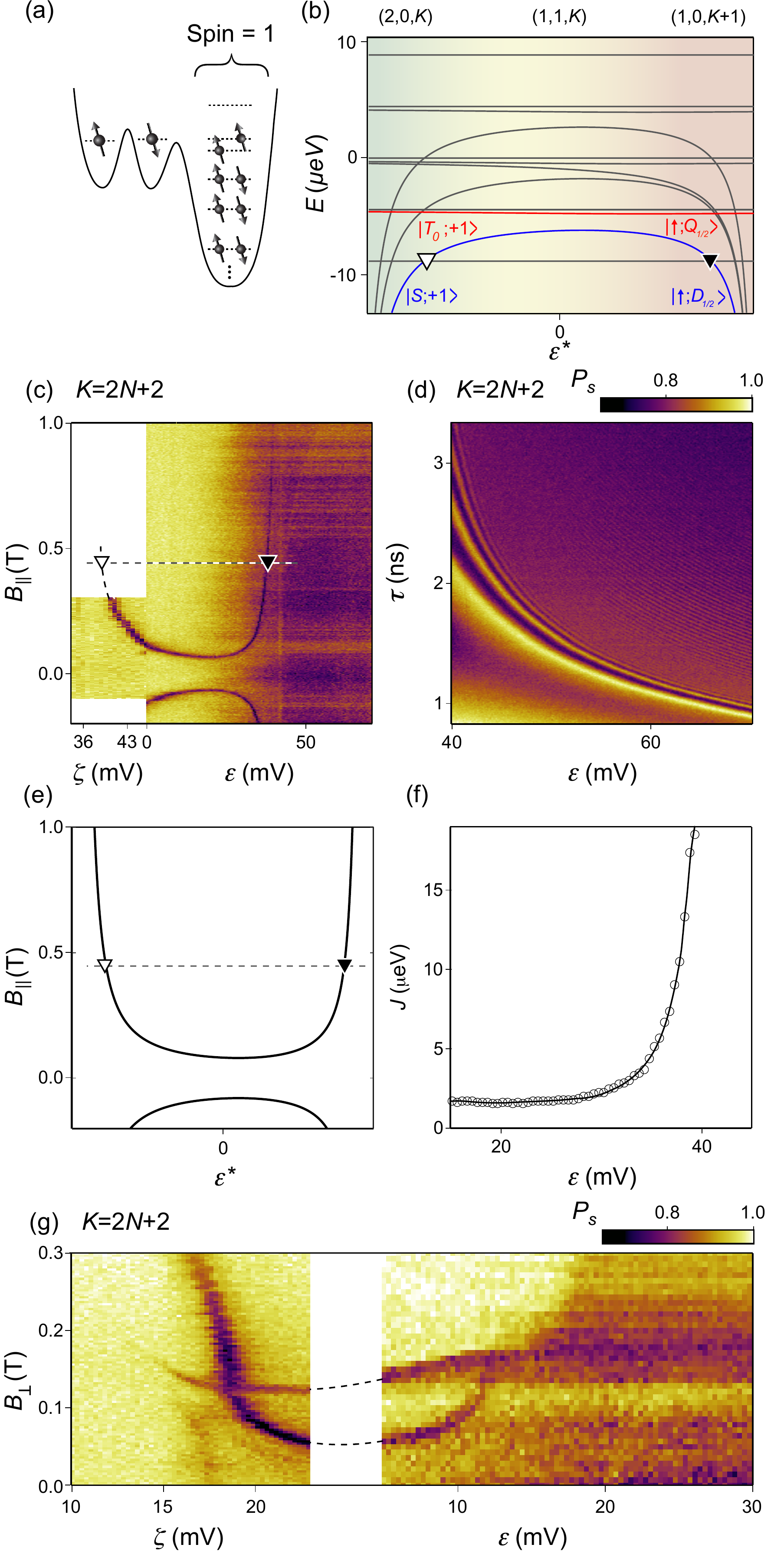}
	\caption{
	(a) Schematic of the odd-occupied multielectron quantum dot with spin-1 ground state, tunnel coupled to the two-electron double quantum dot.
	(b) The inferred energy diagram at the transition between $(2,0,2N \! + \! 2)$, $(1,1,2N \! + \! 2)$ and $(1,0,2N \! + \! 3)$ electronic configurations, for a finite magnetic field. The markers indicate the crossings revealed by the leakage spectroscopy measurement presented in panel (c).
	(d) Time resolved measurement of the exchange oscillations between the $2N \! + \! 2$-occupied spin-1/2 multielectron quantum dot and the neighboring electron.
	(e) Calculated leakage spectrum, extracted from the calculated energy diagram as described in Section \ref{sec:0}.
	(f) Dependence of the exchange energy extracted from the leakage spectroscopy measurement in panel (c).
	(g) Leakage spectroscopy measurement in out-of-plane magnetic field
	}
	\vspace{-20pt}
	\label{fig13}
\end{figure}

Our final case concerns the $2N \! + \! 2$ occupancy of the multielectron quantum dot, which showed different characteristics than the other even occupancies presented. From the behavior of the $2N \! + \! 1$ occupancy (Subsection~\ref{subsec:negative-J}) we concluded that the addition of one electron to the spin-1/2 ground state results in a triplet configuration that has a lower energy than the singlet configuration. Our expectation therefore is that the $2N \! + \! 2$ ground state of the multielectron dot shows spin-1 behavior [Fig.~\ref{fig13}(a)].

Indeed, the leakage spectroscopy data, featuring a prominent U-shaped leakage feature in Fig.~\ref{fig13}(c), is more similar to that of a three-electron triple-dot [Fig.~\ref{fig5}(c)] than to that associated with other even occupations [Figs.~\ref{fig3}(c) and \ref{fig4}]. This points towards the presence of a non-zero spin in the multielectron quantum dot, which we associate with spin 1 in this case. We note that the sharp leakage feature diverges to large $B$ for increasing $\varepsilon$, indicating a positive sign of the exchange interaction (i.e. preferring the low-spin state). Further, measurement of the exchange oscillations clearly show the presence of a coherent exchange interaction [Fig.~\ref{fig13}(d)]. These observations lead us to the following conclusions.  First, the multielectron dot in $2N \! + \! 2$ occupancy carries a non-zero spin.  Second, the exchange interaction with the middle spin has a positive sign, and therefore the transfer of the middle electron to the multielectron dot would result in a reduction of the ground state spin (from spin 1 to  spin 1/2).

Insight into this regime can be gained from the Hubbard model, by calculating the energy spectrum associated with four electrons using Eq.~\ref{eq:spin12Ham}. Figure~\ref{fig13}(b) shows the result for a choice of input parameters that mimics the phenomenology of the multielectron device ($t_1 = 26$~$\mu$eV, $t_2 = 12$~$\mu$eV, $t_{DD} = 30$~$\mu$eV and  $\Delta E=40$~$\mu$eV). 
As before, inspection of this energy diagram allows us to associate spin states with all eigenstates, and infer the expected leakage spectrum from various state crossings. 
In particular, we associate the field-dependent state crossings arising in this energy diagram (marked by a white and black triangle) with the sharp leakage features observed in Fig.~\ref{fig13}(c).

On the left side of the diagram the multielectron dot is decoupled from the double dot by a large negative detuning voltage, and can be viewed as a spin-1 spectator spin. Thus, the eigenstates are tensor products of double dot states ($\ket{S}$, $\ket{T_i}$ with $i=+, -, 0$) and spin-1 states with different spin projections in the direction of the magnetic field ($\ket{0}$ or $\ket{\pm 1}$).
The three states that diverge towards small energies for large negative $\varepsilon^*$ all involve the double-dot singlet state ($\ket{S}$), whereas the other states involve the double-dot triplet states.  Therefore, the state plotted in blue can be viewed as a singlet-like state, whereas the state in red is triplet-like.  

In contrast, for large $\varepsilon^*$, the left dot decouples and becomes a spin-1/2 spectator, while the middle dot (with spin-1/2) hybridizes with the multielectron dot (spin-1) due to tunnel coupling. Accordingly, each eigenstate is the tensor product of a spin-1/2 state and one out of six ``molecular states'' formed by the middle and right dot. In terms of spin, these six ``molecular states'' comprise four quadruplet states ($\ket{Q_{\pm3/2}}$, $\ket{Q_{\pm1/2}}$) with a total spin 3/2, and two doublet states ($\ket{D_{\pm1/2}}$) with a total spin of 1/2 (the subscript in our notation indicates the spin projection in the magnetic field direction). 
For doublet states the middle electron can relocate into the multielectron dot for large positive $\varepsilon^*$. Therefore, all four tensor products that involve doublet states diverge towards low energy in this regime. In contrast, spin-3/2 states within the multielectron dot would be costly in terms of single-particle energies, and hence tensor products that involve the quadruplet states have associated with them a relatively stiff $(1,1,K)$ charge distribution, and transition into $(1,0,K \! + \! 1)$ only for larger $\varepsilon^*$.  

We believe that in the experiment we initialize the triple dot in the $\ket{S;+1}$ state (in analogy to $\ket{S;\uparrow}$ for the spin-1/2 multielectron dot). With increasing detuning this eigenstate continuously changes into $\ket{\uparrow;D_{1/2}}$ [blue line in Fig.~\ref{fig13}(b)]. This change of eigenstates explains the presence of exchange oscillations when pulsing $\varepsilon$ diabatically [Fig.~\ref{fig13}(c)]. Meanwhile, the sharp features in the leakage spectrum correspond to the crossing of this blue-colored state with a fully polarized $\ket{T_+;+1} \equiv \ket{\uparrow; Q_{3/2}}$ state [black and white triangles in Fig.~\ref{fig13}(b),(c)]. Indeed, the leakage spectrum inferred from the calculated energy spectrum confirms this analysis [Fig~\ref{fig13}(e)].

Finally, we present leakage spectroscopy measurements for out-of-plane magnetic field, $B_\perp$ [Fig.~\ref{fig13}(e)]. Curiously, in this case we observe an additional leakage feature with a weaker dependence on detuning. At $\zeta\approx 18$~mV, i.e. near the boundary between $(2,0,2N \! + \! 2)$ and $(1,1,2N \! + \! 2)$, this feature appears to cross the primary feature (which we associate with the $S$-$T_+$ crossing within the double dot) without any sign of interaction. On the other hand, at the transition between $(1,1,2N \! + \! 2)$ and $(1,0,2N \! + \! 3)$ (higher detuning of the multielectron dot) the primary leakage feature ends at exactly the point where the additional feature crosses ($\varepsilon\approx13$~mV, $B_\perp \lesssim 180$~mT). We speculate that the additional feature arises from the strong coupling of multielectron-dot orbitals to the out-of-plane magnetic field, which breaks the near degeneracy between two orbitals~\cite{Hofmann2016} and drives a transition of the ground state spin from 1 ($B_\perp \lesssim 180$~mT) to 0 ($B_\perp \gtrsim 180$~mT). 
This would explain the termination of the primary leakage feature at $\varepsilon\approx13$~mV (since we know from Sec.~\ref{sec:0} that the primary leakage feature is absent when the multielectron quantum dot has a spin-0 ground state) as well as the absence of any interaction near $\zeta\approx 18$~mV (since the double dot in this low detuning is essentially decoupled from the multielectron dot).

\section{Summary and outlook}
\label{sec:summary}

We apply two methods developed for spin qubits to study the spin properties of a mesoscopic multielectron quantum dot, namely spin leakage spectroscopy and measurement of coherent exchange oscillations. Both methods rely on fast gate-voltage pulses, single-shot charge detection, and spin-to-charge conversion, and provide complementary information about the multielectron dot (namely incoherent spin leakage and coherent spin exchange processes). This allows us for the first time to study the spin spectrum associated with the multielectron dot (containing of order 100 electrons) and its dependence on the parity and charge occupation of the multielectron dot.  By studying in detail the interaction of the multielectron dot with a neighboring electron (which we entangle with an unpaired reference electron prior to each pulse cycle), we discover a counterintuitive exchange profile between the neighboring electron and the odd-occupied multielectron quantum dot. Specifically, we observe that the exchange interaction rapidly varies with detuning voltages applied to the multielectron dot, resulting in local maxima and sign changes of the exchange interaction that can be controlled by few-milivolt changes in gate voltages. We also study even occupations of the multielectron dot, including a configuration with spin-1 ground state. We explain our observations using a Hubbard model. Using realistic parameters we show that it qualitatively reproduces the observed diverse behavior of the multielectron dot. The predicted exchange profiles can be classified into four distinct regimes (two of which have been observed in the multielectron device of this work). 

The key conclusion of this work is that a multielectron quantum dot possesses properties that may be exploited as a mediator of exchange interactions for spin qubit applications. We observe a spin-0 ground state, most desirable for long-range exchange coupling, for 3 out of 4 of the studied even occupancies.  This should provide sufficient reliability for use in a scalable quantum dot system -- accidental spin-1 ground states of the multielectron mediator can be avoided by adding or removing two electrons. A first demonstration towards long-range exchange coupling was recently demonstrated in the same device~\cite{JB_mediated_exchange}.

Several other findings may also benefit spin qubit applications. 
First, the appearance of extrema in the observed exchange profiles may be suitable to increase gate fidelities, by reducing the sensitivity of exchange oscillations to charge noise~\cite{Martins2016,Reed2016}. 
Second, access to both signs of the exchange lifts constraints for the construction of dynamically decoupled gates. Previous theoretical work assumed an exchange coupling that can only assume zero and positive values, and the resulting gate sequences can be quite complex~\cite{Wang2012,Wang2014}. 
Third, since the large quantum dot is characterized by a reduced level spacing, it may be possible to define a singlet-triplet rotating frame on the multielectron dot that is charge-noise insensitive (as demonstrated in Ref.~\onlinecite{Dial2013}) but doesn't suffer from impractically high operating frequencies (analogous to the quantum dot hybrid qubit~\cite{Kim2014,Cao2016}). 
Fourth, a larger size of the multielectron quantum dot implies a reduction of the Overhauser field experienced by the electrons, and therefore a reduced dephasing rate~\cite{Taylor2007}. Fifth, the addition of the single electron to the spin-0 quantum dot preserves the spin of the electron. Therefore, it should be possible to subsequently eject this electron to another quantum dot, yielding a method for coherent shuttling of spin states between distant quantum dots~\cite{Fujita2017}.

From a fundamental physics point of view, several aspects of the multielectron quantum dot could be studied further. For example, the distribution of level spacings and the strength of the spin correlation energy are likely characterized by mesoscopic fluctuations, and were not studied here. Their dependence on the dot size is of fundamental and practical importance. Another curiosity is that for all three spin-1/2 ground states for which the quality of data allowed a full analysis (i.e. $K= 2N\! - \! 3,2N\! - \! 1, 2N\! + \! 1$) we observe extrema in the exchange strength, both for positive and negative exchange strengths. This may hint towards a correlation between the level spacing and the ratio of the tunnel couplings.

Finally, for the first time we apply leakage spectroscopy and exchange oscillations measurements of a spin qubit to study the spectrum of more complicated, largely unknown object (the multielectron dot). The same principle could be applied to study numerous other systems and poorly understood phenomena. Examples include quantum dots coupled
to quantum Hall or fractional quantum Hall edge states \cite{Yang2015,Kiyama2015}, or to hybrid super-semiconducting quantum dots such as Majorana islands~\cite{Gharavi2016,Deng2016}. A scanning probe version of this technique, in which a spin qubit is scanned over surfaces as in a scanning tunneling microscope, would open the study of exchange interactions to an even larger class of quantum materials.

\section*{Acknowledgements}

We thank E. Barnes for helpful discussions.
This work was supported by the Army Research Office, the Innovation Fund Denmark, the Villum Foundation, and the Danish National Research Foundation. Work at Sydney was supported by the ARC via the Centre of Excellence in Engineered Quantum Systems (EQuS), project number CE110001013. 
Work at Purdue was supported by the U.S. Department of Energy, Office of Basic Energy Sciences, Division of Materials, Sciences and Engineering under Award No. DE-SC0006671. Additional support from Nokia Bell Labs for the GaAs MBE effort is also gratefully acknowledged.

\medskip

\appendix

\section{Parameters used to calculate the presented energy spectra}
\label{app:parameters}

The Hamiltonians \eqref{eq:spin0Ham} and \eqref{eq:spin12Ham} contain parameters that were not measured directly, but which were estimated based on realistic experimental assumptions. These include the on- and off-site Coulomb interaction energies $U_i$ and $K_{ij}$, and the spin correlation energy $\xi$. To reduce the number of parameters in our modeling, we also fix certain combinations of single particle energies, namely $\bar{\varepsilon} = (\varepsilon_L + \varepsilon_M +\varepsilon_{R(1)})/3$ and $\varepsilon_M=\varepsilon_M-(\varepsilon_L + \varepsilon_{R(1)})/2$. All energy diagrams presented in this paper were calculated using identical sets of parameters, summarized in table~\ref{tab2}, but differed in the assumed number of occupied orbitals appropriate for the different charge occupations of the multielectron dot.

\begin{table}[!h]
	\caption{ Summary of parameters used in the Hubbard model for all presented simulations. }
	\label{tab2}
\begin{ruledtabular}
\begin{tabular}{cc}
\textbf{Fixed parameters} & \textbf{Value (meV)} \\ \hline
$U_L = U_M$ & 5 \\
$U_R = U_{R1} = U_{R2} = K_{R1,R2}$ & 1 \\
$K_{LM} = K_{MR} = K_{MR1} = K_{MR2}$ & 0.1 \\
$K_{LR} = K_{LR1} = K_{LR2}$ & 0.02 \\
$\xi$ & 0.1 \\
$\varepsilon_M$ & 2 \\
$\bar{\varepsilon}$ & 0
\end{tabular} 
\end{ruledtabular}
\end{table}


\begin{thebibliography}{10}
\expandafter\ifx\csname url\endcsname\relax
  \def\url#1{\texttt{#1}}\fi
\expandafter\ifx\csname urlprefix\endcsname\relax\def\urlprefix{URL }\fi
\providecommand{\bibinfo}[2]{#2}
\providecommand{\eprint}[2][]{\url{#2}}

\bibitem{Nowack2011}

\newblock \bibinfo{author}{Nowack, K.~C.} \emph{et~al.}
\newblock \emph{\bibinfo{journal}{Science}} \textbf{\bibinfo{volume}{333}},
  \bibinfo{pages}{1269--1272} (\bibinfo{year}{2011}).

\bibitem{Shulman2012}

\newblock \bibinfo{author}{Shulman, M.~D.}, \bibinfo{author}{Dial, O.~E.},
  \bibinfo{author}{Harvey, S.~P.}, \bibinfo{author}{Bluhm, H.},
  \bibinfo{author}{Umansky, V.} \& \bibinfo{author}{Yacoby, A.}
\newblock \emph{\bibinfo{journal}{Science}} \textbf{\bibinfo{volume}{336}},
  \bibinfo{pages}{202--205} (\bibinfo{year}{2012}).

\bibitem{Gaudreau2011}

\newblock \bibinfo{author}{Gaudreau, L.} \emph{et~al.}
\newblock \emph{\bibinfo{journal}{Nature Physics}}
  \textbf{\bibinfo{volume}{8}}, \bibinfo{pages}{54--58} (\bibinfo{year}{2011}).

\bibitem{Studenikin2012}

\newblock \bibinfo{author}{Studenikin, S.~A.} \emph{et~al.}
\newblock \emph{\bibinfo{journal}{Physical Review Letters}}
  \textbf{\bibinfo{volume}{108}}, \bibinfo{pages}{226802}
  (\bibinfo{year}{2012}).

\bibitem{Cao2016}

\newblock \bibinfo{author}{Cao, G.} \emph{et~al.}
\newblock \emph{\bibinfo{journal}{Physical Review Letters}}
  \textbf{\bibinfo{volume}{116}}, \bibinfo{pages}{086801}
  (\bibinfo{year}{2016}).

\bibitem{Malinowski2017}

\newblock \bibinfo{author}{Malinowski, F.~K.} \emph{et~al.}
\newblock \emph{\bibinfo{journal}{Nature Nanotechnology}}
  \textbf{\bibinfo{volume}{12}}, \bibinfo{pages}{16--20}
  (\bibinfo{year}{2017}).

\bibitem{Bertrand2016}

\newblock \bibinfo{author}{Bertrand, B.} \emph{et~al.}
\newblock \emph{\bibinfo{journal}{Nature Nanotechnology}}
  \bibinfo{pages}{672--676} (\bibinfo{year}{2016}).

\bibitem{Maune2012}

\newblock \bibinfo{author}{Maune, B.~M.} \emph{et~al.}
\newblock \emph{\bibinfo{journal}{Nature}} \textbf{\bibinfo{volume}{481}},
  \bibinfo{pages}{344--347} (\bibinfo{year}{2012}).

\bibitem{Kim2014}

\newblock \bibinfo{author}{Kim, D.} \emph{et~al.}
\newblock \emph{\bibinfo{journal}{Nature}} \textbf{\bibinfo{volume}{511}},
  \bibinfo{pages}{70} (\bibinfo{year}{2014}).

\bibitem{Eng2015}

\newblock \bibinfo{author}{Eng, K.} \emph{et~al.}
\newblock \emph{\bibinfo{journal}{Science Advances}}
  \textbf{\bibinfo{volume}{1}}, \bibinfo{pages}{1500214}
  (\bibinfo{year}{2015}).

\bibitem{Kawakami2016}

\newblock \bibinfo{author}{Kawakami, E.} \emph{et~al.}
\newblock \emph{\bibinfo{journal}{Proceedings of the National Academy of
  Sciences}} \textbf{\bibinfo{volume}{113}}, \bibinfo{pages}{11738--11743}
  (\bibinfo{year}{2016}).

\bibitem{Takeda2016}

\newblock \bibinfo{author}{Takeda, K.} \emph{et~al.}
\newblock \emph{\bibinfo{journal}{Science Advances}}
  \textbf{\bibinfo{volume}{2}}, \bibinfo{pages}{e1600694}
  (\bibinfo{year}{2016}).

\bibitem{Veldhorst2014}

\newblock \bibinfo{author}{Veldhorst, M.} \emph{et~al.}
\newblock \emph{\bibinfo{journal}{Nature Nanotechnology}}
  \textbf{\bibinfo{volume}{9}}, \bibinfo{pages}{981--985}
  (\bibinfo{year}{2014}).

\bibitem{Maurand2016}

\newblock \bibinfo{author}{Maurand, R.} \emph{et~al.}
\newblock \emph{\bibinfo{journal}{Nature Communications}}
  \textbf{\bibinfo{volume}{7}}, \bibinfo{pages}{13575} (\bibinfo{year}{2016}).

\bibitem{Pla2012}

\newblock \bibinfo{author}{Pla, J.~J.} \emph{et~al.}
\newblock \emph{\bibinfo{journal}{Nature}} \textbf{\bibinfo{volume}{489}},
  \bibinfo{pages}{541--545} (\bibinfo{year}{2012}).

\bibitem{Muhonen2014}

\newblock \bibinfo{author}{Muhonen, J.~T.} \emph{et~al.}
\newblock \emph{\bibinfo{journal}{Nature Nanotechnology}}
  \textbf{\bibinfo{volume}{9}}, \bibinfo{pages}{986--991}
  (\bibinfo{year}{2014}).

\bibitem{Foletti2009}

\newblock \bibinfo{author}{Foletti, S.}, \bibinfo{author}{Bluhm, H.},
  \bibinfo{author}{Mahalu, D.}, \bibinfo{author}{Umansky, V.} \&
  \bibinfo{author}{Yacoby, A.}
\newblock \emph{\bibinfo{journal}{Nature Physics}}
  \textbf{\bibinfo{volume}{5}}, \bibinfo{pages}{903--908}
  (\bibinfo{year}{2009}).

\bibitem{Cerfontaine2016}

\newblock \bibinfo{author}{Cerfontaine, P.}, \bibinfo{author}{Botzem, T.},
  \bibinfo{author}{Humpohl, S.~S.}, \bibinfo{author}{Schuh, D.},
  \bibinfo{author}{Bougeard, D.} \& \bibinfo{author}{Bluhm, H.}
\newblock \emph{\bibinfo{journal}{arXiv}} \bibinfo{pages}{arXiv: 1606.01897}
  (\bibinfo{year}{2016}).

\bibitem{Medford2013a}

\newblock \bibinfo{author}{Medford, J.}, \bibinfo{author}{Beil, J.},
  \bibinfo{author}{Taylor, J.}, \bibinfo{author}{Rashba, E.~I.},
  \bibinfo{author}{Lu, H.}, \bibinfo{author}{Gossard, A.~C.} \&
  \bibinfo{author}{Marcus, C.~M.}
\newblock \emph{\bibinfo{journal}{Physical Review Letters}}
  \textbf{\bibinfo{volume}{111}}, \bibinfo{pages}{050501}
  (\bibinfo{year}{2013}).

\bibitem{Medford2013}

\newblock \bibinfo{author}{Medford, J.} \emph{et~al.}
\newblock \emph{\bibinfo{journal}{Nature Nanotechnology}}
  \textbf{\bibinfo{volume}{8}}, \bibinfo{pages}{654--659}
  (\bibinfo{year}{2013}).

\bibitem{Nichol2017}

\newblock \bibinfo{author}{Nichol, J.~M.}, \bibinfo{author}{Orona, L.~A.},
  \bibinfo{author}{Harvey, S.~P.}, \bibinfo{author}{Fallahi, S.},
  \bibinfo{author}{Gardner, G.~C.}, \bibinfo{author}{Manfra, M.~J.} \&
  \bibinfo{author}{Yacoby, A.}
\newblock \emph{\bibinfo{journal}{npj Quantum Information}}
  \textbf{\bibinfo{volume}{3}}, \bibinfo{pages}{3} (\bibinfo{year}{2017}).

\bibitem{Veldhorst2015}

\newblock \bibinfo{author}{Veldhorst, M.} \emph{et~al.}
\newblock \emph{\bibinfo{journal}{Nature}} \textbf{\bibinfo{volume}{526}},
  \bibinfo{pages}{410--414} (\bibinfo{year}{2015}).

\bibitem{Watson2017}

\newblock \bibinfo{author}{Watson, T.~F.} \emph{et~al.}
\newblock \emph{\bibinfo{journal}{arXiv}} \bibinfo{pages}{arXiv: 1708.04214}
  (\bibinfo{year}{2017}).

\bibitem{Martins2016}

\newblock \bibinfo{author}{Martins, F.} \emph{et~al.}
\newblock \emph{\bibinfo{journal}{Physical Review Letters}}
  \textbf{\bibinfo{volume}{116}}, \bibinfo{pages}{116801}
  (\bibinfo{year}{2016}).

\bibitem{Reed2016}

\newblock \bibinfo{author}{Reed, M.~D.} \emph{et~al.}
\newblock \emph{\bibinfo{journal}{Physical Review Letters}}
  \textbf{\bibinfo{volume}{116}}, \bibinfo{pages}{110402}
  (\bibinfo{year}{2016}).

\bibitem{Taylor2005a}

\newblock \bibinfo{author}{Taylor, J.}, \bibinfo{author}{Engel, H.-A.},
  \bibinfo{author}{D{\"{u}}r, W.}, \bibinfo{author}{Yacoby, A.},
  \bibinfo{author}{Marcus, C.~M.}, \bibinfo{author}{Zoller, P.} \&
  \bibinfo{author}{Lukin, M.~D.}
\newblock \emph{\bibinfo{journal}{Nature Physics}}
  \textbf{\bibinfo{volume}{1}}, \bibinfo{pages}{177--183}
  (\bibinfo{year}{2005}).

\bibitem{Wang2015}

\newblock \bibinfo{author}{Klochan, O.} \emph{et~al.}
\newblock \emph{\bibinfo{journal}{Nanotechnology}}
  \textbf{\bibinfo{volume}{494}}, \bibinfo{pages}{155329}
  (\bibinfo{year}{2015}).

\bibitem{Zajac2016}

\newblock \bibinfo{author}{Zajac, D.~M.}, \bibinfo{author}{Hazard, T.~M.},
  \bibinfo{author}{Mi, X.}, \bibinfo{author}{Nielsen, E.} \&
  \bibinfo{author}{Petta, J.~R.}
\newblock \emph{\bibinfo{journal}{Physical Review Applied}}
  \textbf{\bibinfo{volume}{6}}, \bibinfo{pages}{054013} (\bibinfo{year}{2016}).

\bibitem{Burkard2006}

\newblock \bibinfo{author}{Burkard, G.} \& \bibinfo{author}{Imamoglu, A.}
\newblock \emph{\bibinfo{journal}{Physical Review B}}
  \textbf{\bibinfo{volume}{74}}, \bibinfo{pages}{041307(R)}
  (\bibinfo{year}{2006}).

\bibitem{Liu2014}

\newblock \bibinfo{author}{Liu, Y.-Y.}, \bibinfo{author}{Petersson, K.~D.},
  \bibinfo{author}{Stehlik, J.}, \bibinfo{author}{Taylor, J.} \&
  \bibinfo{author}{Petta, J.~R.}
\newblock \emph{\bibinfo{journal}{Physical Review Letters}}
  \textbf{\bibinfo{volume}{113}}, \bibinfo{pages}{036801}
  (\bibinfo{year}{2014}).

\bibitem{Viennot2015}

\newblock \bibinfo{author}{Viennot, J.~J.}, \bibinfo{author}{Dartiailh, M.~C.},
  \bibinfo{author}{Cottet, A.} \& \bibinfo{author}{Kontos, T.}
\newblock \emph{\bibinfo{journal}{Science}} \textbf{\bibinfo{volume}{349}},
  \bibinfo{pages}{408--411} (\bibinfo{year}{2015}).

\bibitem{Mi2016}

\newblock \bibinfo{author}{Mi, X.}, \bibinfo{author}{Cady, J.~V.},
  \bibinfo{author}{Zajac, D.~M.}, \bibinfo{author}{Deelman, P.~W.} \&
  \bibinfo{author}{Petta, J.~R.}
\newblock \emph{\bibinfo{journal}{Science}} \textbf{\bibinfo{volume}{355}},
  \bibinfo{pages}{156--158} (\bibinfo{year}{2017}).

\bibitem{Russ2015a}

\newblock \bibinfo{author}{Russ, M.} \& \bibinfo{author}{Burkard, G.}
\newblock \emph{\bibinfo{journal}{Physical Review B}}
  \textbf{\bibinfo{volume}{92}}, \bibinfo{pages}{205412}
  (\bibinfo{year}{2015}).

\bibitem{Srinivasa2016}

\newblock \bibinfo{author}{Srinivasa, V.}, \bibinfo{author}{Taylor, J.} \&
  \bibinfo{author}{Tahan, C.}
\newblock \emph{\bibinfo{journal}{Physical Review B}}
  \textbf{\bibinfo{volume}{94}}, \bibinfo{pages}{205421}
  (\bibinfo{year}{2016}).

\bibitem{Srinivasa2015}

\newblock \bibinfo{author}{Srinivasa, V.}, \bibinfo{author}{Xu, H.} \&
  \bibinfo{author}{Taylor, J.}
\newblock \emph{\bibinfo{journal}{Physical Review Letters}}
  \textbf{\bibinfo{volume}{114}}, \bibinfo{pages}{226803}
  (\bibinfo{year}{2015}).

\bibitem{Mehl2014a}

\newblock \bibinfo{author}{Mehl, S.}, \bibinfo{author}{Bluhm, H.} \&
  \bibinfo{author}{DiVincenzo, D.~P.}
\newblock \emph{\bibinfo{journal}{Physical Review B}}
  \textbf{\bibinfo{volume}{90}}, \bibinfo{pages}{045404}
  (\bibinfo{year}{2014}).

\bibitem{Baart2017}

\newblock \bibinfo{author}{Baart, T.~A.}, \bibinfo{author}{Fujita, T.},
  \bibinfo{author}{Reichl, C.}, \bibinfo{author}{Wegscheider, W.} \&
  \bibinfo{author}{Vandersypen, L. M.~K.}
\newblock \emph{\bibinfo{journal}{Nature Nanotechnology}}
  \textbf{\bibinfo{volume}{12}}, \bibinfo{pages}{26--30}
  (\bibinfo{year}{2017}).

\bibitem{JB_mediated_exchange}

\newblock \bibinfo{author}{Malinowski, F.~K.} \emph{et~al.} \bibinfo{pages}{(in
  preparation)}.

\bibitem{Folk1996}

\newblock \bibinfo{author}{Folk, J.~A.} \emph{et~al.}
\newblock \emph{\bibinfo{journal}{Physical Review Letters}}
  \textbf{\bibinfo{volume}{76}}, \bibinfo{pages}{1699--1702}
  (\bibinfo{year}{1996}).

\bibitem{Kouwenhoven1997}

\newblock \bibinfo{author}{Kouwenhoven, L.~P.}, \bibinfo{author}{Oosterkamp,
  T.~H.}, \bibinfo{author}{Danoesastro, M. W.~S.}, \bibinfo{author}{Eto, M.},
  \bibinfo{author}{Austing, D.~G.}, \bibinfo{author}{Honda, T.} \&
  \bibinfo{author}{Tarucha, S.}
\newblock \emph{\bibinfo{journal}{Science}} \textbf{\bibinfo{volume}{278}},
  \bibinfo{pages}{185--205} (\bibinfo{year}{1997}).

\bibitem{Stewart1997}

\newblock \bibinfo{author}{Stewart, D.~R.}, \bibinfo{author}{Sprinzak, D.},
  \bibinfo{author}{Marcus, C.~M.}, \bibinfo{author}{Duruo, C.~I.} \&
  \bibinfo{author}{Jr, J. S.~H.}
\newblock \emph{\bibinfo{journal}{Science}} \textbf{\bibinfo{volume}{278}},
  \bibinfo{pages}{1784--1788} (\bibinfo{year}{1997}).

\bibitem{Folk2001}

\newblock \bibinfo{author}{Folk, J.~A.}, \bibinfo{author}{Marcus, C.~M.},
  \bibinfo{author}{Berkovits, R.}, \bibinfo{author}{Kurland, I.~L.},
  \bibinfo{author}{Aleiner, I.~L.} \& \bibinfo{author}{Altshuler, B.~L.}
\newblock \emph{\bibinfo{journal}{Physica Scripta}}
  \textbf{\bibinfo{volume}{T90}}, \bibinfo{pages}{26--33}
  (\bibinfo{year}{2001}).

\bibitem{Alhassid2000}

\newblock \bibinfo{author}{Alhassid, Y.}
\newblock \emph{\bibinfo{journal}{Reviews of Modern Physics}}
  \textbf{\bibinfo{volume}{72}}, \bibinfo{pages}{895--968}
  (\bibinfo{year}{2000}).

\bibitem{Negative-J}

\newblock \bibinfo{author}{Martins, F.} \emph{et~al.} \bibinfo{pages}{(in
  preparation)}.

\bibitem{Croot2017}

\newblock \bibinfo{author}{Croot, X.~G.}, \bibinfo{author}{Pauka, S.~J.},
  \bibinfo{author}{Watson, J.~D.}, \bibinfo{author}{Gardner, G.~C.},
  \bibinfo{author}{Fallahi, S.}, \bibinfo{author}{Manfra, M.~J.} \&
  \bibinfo{author}{Reilly, D.~J.}
\newblock \emph{\bibinfo{journal}{arXiv}} \bibinfo{pages}{arXiv: 1707.06479}
  (\bibinfo{year}{2017}).

\bibitem{Hu2001}

\newblock \bibinfo{author}{Hu, X.} \& \bibinfo{author}{Sarma, S.~D.}
\newblock \emph{\bibinfo{journal}{Physical Review A}}
  \textbf{\bibinfo{volume}{64}}, \bibinfo{pages}{042312}
  (\bibinfo{year}{2001}).

\bibitem{Buizert2008}

\newblock \bibinfo{author}{Buizert, C.} \emph{et~al.}
\newblock \emph{\bibinfo{journal}{Physical Review Letters}}
  \textbf{\bibinfo{volume}{101}}, \bibinfo{pages}{226603}
  (\bibinfo{year}{2008}).

\bibitem{Yoneda2014}

\newblock \bibinfo{author}{Yoneda, J.} \emph{et~al.}
\newblock \emph{\bibinfo{journal}{Physical Review Letters}}
  \textbf{\bibinfo{volume}{113}}, \bibinfo{pages}{267601}
  (\bibinfo{year}{2014}).

\bibitem{Petta2005}

\newblock \bibinfo{author}{Petta, J.~R.} \emph{et~al.}
\newblock \emph{\bibinfo{journal}{Science}} \textbf{\bibinfo{volume}{309}},
  \bibinfo{pages}{2180--2184} (\bibinfo{year}{2005}).

\bibitem{Reilly2008}

\newblock \bibinfo{author}{Reilly, D.~J.}, \bibinfo{author}{Taylor, J.},
  \bibinfo{author}{Laird, E.~A.}, \bibinfo{author}{Petta, J.~R.},
  \bibinfo{author}{Marcus, C.~M.}, \bibinfo{author}{Hanson, M.~P.} \&
  \bibinfo{author}{Gossard, A.~C.}
\newblock \emph{\bibinfo{journal}{Physical Review Letters}}
  \textbf{\bibinfo{volume}{101}}, \bibinfo{pages}{236803}
  (\bibinfo{year}{2008}).

\bibitem{Bluhm2010}

\newblock \bibinfo{author}{Bluhm, H.}, \bibinfo{author}{Foletti, S.},
  \bibinfo{author}{Mahalu, D.}, \bibinfo{author}{Umansky, V.} \&
  \bibinfo{author}{Yacoby, A.}
\newblock \emph{\bibinfo{journal}{Physical Review Letters}}
  \textbf{\bibinfo{volume}{105}}, \bibinfo{pages}{216803}
  (\bibinfo{year}{2010}).

\bibitem{Datta1997}
\bibinfo{author}{Datta, S.}
\newblock \emph{\bibinfo{title}{Electronic transport in mesoscopic systems}}
  (\bibinfo{publisher}{Cambridge university press}, \bibinfo{year}{1997}).

\bibitem{Brouwer1999}

\newblock \bibinfo{author}{Brouwer, P.~W.}, \bibinfo{author}{Oreg, Y.} \&
  \bibinfo{author}{Halperin, B.~I.}
\newblock \emph{\bibinfo{journal}{Physical Review B}}
  \textbf{\bibinfo{volume}{60}}, \bibinfo{pages}{977--980}
  (\bibinfo{year}{1999}).

\bibitem{Kurland2000}

\newblock \bibinfo{author}{Kurland, I.~L.}, \bibinfo{author}{Kurland, I.~L.},
  \bibinfo{author}{Aleiner, I.~L.}, \bibinfo{author}{Aleiner, I.~L.},
  \bibinfo{author}{Altshuler, B.~L.}, \bibinfo{author}{Altshuler, B.~L.} \&
  \bibinfo{author}{Altshuler, B.~L.}
\newblock \emph{\bibinfo{journal}{Physical Review B}}
  \textbf{\bibinfo{volume}{62}}, \bibinfo{pages}{14886--14897}
  (\bibinfo{year}{2000}).

\bibitem{Baranger2000}

\newblock \bibinfo{author}{Baranger, H.~U.}, \bibinfo{author}{Ullmo, D.} \&
  \bibinfo{author}{Glazman, L.~I.}
\newblock \emph{\bibinfo{journal}{Physical Review B}}
  \textbf{\bibinfo{volume}{61}}, \bibinfo{pages}{R2425} (\bibinfo{year}{2000}).

\bibitem{Blanter1997}

\newblock \bibinfo{author}{Blanter, Y.~M.}, \bibinfo{author}{Mirlin, A.~D.} \&
  \bibinfo{author}{Muzykantskii, B.~A.}
\newblock \emph{\bibinfo{journal}{Physical Review Letters}}
  \textbf{\bibinfo{volume}{78}}, \bibinfo{pages}{2449--2452}
  (\bibinfo{year}{1997}).

\bibitem{Hirose2002}

\newblock \bibinfo{author}{Hirose, K.} \& \bibinfo{author}{Wingreen, N.~S.}
\newblock \emph{\bibinfo{journal}{Physical Review B}}
  \textbf{\bibinfo{volume}{65}}, \bibinfo{pages}{193305}
  (\bibinfo{year}{2002}).

\bibitem{Sivan1996}

\newblock \bibinfo{author}{Sivan, U.}, \bibinfo{author}{Berkovits, R.},
  \bibinfo{author}{Aloni, Y.}, \bibinfo{author}{Prus, O.} \&
  \bibinfo{author}{Auerbach, A.}
\newblock \emph{\bibinfo{journal}{Physical Review Letters}}
  \textbf{\bibinfo{volume}{77}}, \bibinfo{pages}{1123--1126}
  (\bibinfo{year}{1996}).

\bibitem{Gorokhov2004}

\newblock \bibinfo{author}{Gorokhov, D.~A.} \& \bibinfo{author}{Brouwer, P.~W.}
\newblock \emph{\bibinfo{journal}{Physical Review B}}
  \textbf{\bibinfo{volume}{69}}, \bibinfo{pages}{155417}
  (\bibinfo{year}{2004}).

\bibitem{Andreev1998}

\newblock \bibinfo{author}{Andreev, A.~V.} \& \bibinfo{author}{Kamenev, A.}
\newblock \emph{\bibinfo{journal}{Physical Review Letters}}
  \textbf{\bibinfo{volume}{81}}, \bibinfo{pages}{3199 --3202}
  (\bibinfo{year}{1998}).

\bibitem{Lindemann2002}

\newblock \bibinfo{author}{Lindemann, S.}, \bibinfo{author}{Ihn, T.},
  \bibinfo{author}{Heinzel, T.}, \bibinfo{author}{Zwerger, W.},
  \bibinfo{author}{Ensslin, K.}, \bibinfo{author}{Maranowski, K.} \&
  \bibinfo{author}{Gossard, A.~C.}
\newblock \emph{\bibinfo{journal}{Physical Review B}}
  \textbf{\bibinfo{volume}{66}}, \bibinfo{pages}{195314}
  (\bibinfo{year}{2002}).

\bibitem{Barthel2009}

\newblock \bibinfo{author}{Barthel, C.}, \bibinfo{author}{Reilly, D.~J.},
  \bibinfo{author}{Marcus, C.~M.}, \bibinfo{author}{Hanson, M.~P.} \&
  \bibinfo{author}{Gossard, A.~C.}
\newblock \emph{\bibinfo{journal}{Physical Review Letters}}
  \textbf{\bibinfo{volume}{103}}, \bibinfo{pages}{160503}
  (\bibinfo{year}{2009}).

\bibitem{Barthel2012}

\newblock \bibinfo{author}{Barthel, C.}, \bibinfo{author}{Medford, J.},
  \bibinfo{author}{Bluhm, H.}, \bibinfo{author}{Yacoby, A.},
  \bibinfo{author}{Marcus, C.~M.}, \bibinfo{author}{Hanson, M.~P.} \&
  \bibinfo{author}{Gossard, A.~C.}
\newblock \emph{\bibinfo{journal}{Physical Review B}}
  \textbf{\bibinfo{volume}{85}}, \bibinfo{pages}{035306}
  (\bibinfo{year}{2012}).

\bibitem{Shulman2014}

\newblock \bibinfo{author}{Shulman, M.~D.}, \bibinfo{author}{Harvey, S.~P.},
  \bibinfo{author}{Nichol, J.~M.}, \bibinfo{author}{Bartlett, S.~D.},
  \bibinfo{author}{Doherty, A.~C.}, \bibinfo{author}{Umansky, V.} \&
  \bibinfo{author}{Yacoby, A.}
\newblock \emph{\bibinfo{journal}{Nature Communications}}
  \textbf{\bibinfo{volume}{5}}, \bibinfo{pages}{5156} (\bibinfo{year}{2014}).

\bibitem{Delbecq2016}

\newblock \bibinfo{author}{Delbecq, M.~R.} \emph{et~al.}
\newblock \emph{\bibinfo{journal}{Physical Review Letters}}
  \textbf{\bibinfo{volume}{116}}, \bibinfo{pages}{046802}
  (\bibinfo{year}{2016}).

\bibitem{Malinowski2017a}

\newblock \bibinfo{author}{Malinowski, F.~K.} \emph{et~al.}
\newblock \emph{\bibinfo{journal}{Physical Review Letters}}
  \textbf{\bibinfo{volume}{118}}, \bibinfo{pages}{177702}
  (\bibinfo{year}{2017}).

\bibitem{Laird2010}

\newblock \bibinfo{author}{Laird, E.~A.}, \bibinfo{author}{Taylor, J.},
  \bibinfo{author}{DiVincenzo, D.~P.}, \bibinfo{author}{Marcus, C.~M.},
  \bibinfo{author}{Hanson, M.~P.} \& \bibinfo{author}{Gossard, A.~C.}
\newblock \emph{\bibinfo{journal}{Physical Review B}}
  \textbf{\bibinfo{volume}{82}}, \bibinfo{pages}{075403}
  (\bibinfo{year}{2010}).

\bibitem{Poulin-Lamarre2015}

\newblock \bibinfo{author}{Poulin-Lamarre, G.} \emph{et~al.}
\newblock \emph{\bibinfo{journal}{Physical Review B}}
  \textbf{\bibinfo{volume}{91}}, \bibinfo{pages}{125417}
  (\bibinfo{year}{2015}).

\bibitem{Dial2013}

\newblock \bibinfo{author}{Dial, O.~E.}, \bibinfo{author}{Shulman, M.~D.},
  \bibinfo{author}{Harvey, S.~P.}, \bibinfo{author}{Bluhm, H.},
  \bibinfo{author}{Umansky, V.} \& \bibinfo{author}{Yacoby, A.}
\newblock \emph{\bibinfo{journal}{Physical Review Letters}}
  \textbf{\bibinfo{volume}{110}}, \bibinfo{pages}{146804}
  (\bibinfo{year}{2013}).

\bibitem{Hofmann2016}

\newblock \bibinfo{author}{Hofmann, A.} \emph{et~al.}
\newblock \emph{\bibinfo{journal}{Physical Review Letters}}
  \textbf{\bibinfo{volume}{117}}, \bibinfo{pages}{206803}
  (\bibinfo{year}{2016}).

\bibitem{Wang2012}

\newblock \bibinfo{author}{Wang, X.}, \bibinfo{author}{Bishop, L.~S.},
  \bibinfo{author}{Kestner, J.~P.}, \bibinfo{author}{Barnes, E.},
  \bibinfo{author}{Sun, K.} \& \bibinfo{author}{{Das Sarma}, S.}
\newblock \emph{\bibinfo{journal}{Nature Communications}}
  \textbf{\bibinfo{volume}{3}}, \bibinfo{pages}{997} (\bibinfo{year}{2012}).

\bibitem{Wang2014}

\newblock \bibinfo{author}{Wang, X.}, \bibinfo{author}{Bishop, L.~S.},
  \bibinfo{author}{Barnes, E.}, \bibinfo{author}{Kestner, J.~P.} \&
  \bibinfo{author}{{Das Sarma}, S.}
\newblock \emph{\bibinfo{journal}{Physical Review A}}
  \textbf{\bibinfo{volume}{89}}, \bibinfo{pages}{022310}
  (\bibinfo{year}{2014}).

\bibitem{Taylor2007}

\newblock \bibinfo{author}{Taylor, J.}, \bibinfo{author}{Petta, J.~R.},
  \bibinfo{author}{Johnson, A.~C.}, \bibinfo{author}{Yacoby, A.},
  \bibinfo{author}{Marcus, C.~M.} \& \bibinfo{author}{Lukin, M.~D.}
\newblock \emph{\bibinfo{journal}{Physical Review B}}
  \textbf{\bibinfo{volume}{76}}, \bibinfo{pages}{035315}
  (\bibinfo{year}{2007}).

\bibitem{Fujita2017}

\newblock \bibinfo{author}{Fujita, T.}, \bibinfo{author}{Baart, T.~A.},
  \bibinfo{author}{Reichl, C.}, \bibinfo{author}{Wegscheider, W.} \&
  \bibinfo{author}{Vandersypen, L. M.~K.}
\newblock \emph{\bibinfo{journal}{arXiv}} \bibinfo{pages}{arXiv: 1701.00815}
  (\bibinfo{year}{2017}).

\bibitem{Yang2015}

\newblock \bibinfo{author}{Yang, G.}, \bibinfo{author}{Hsu, C.-H.~H.},
  \bibinfo{author}{Stano, P.}, \bibinfo{author}{Klinovaja, J.} \&
  \bibinfo{author}{Loss, D.}
\newblock \emph{\bibinfo{journal}{Physical Review B}}
  \textbf{\bibinfo{volume}{93}}, \bibinfo{pages}{075301}
  (\bibinfo{year}{2015}).

\bibitem{Kiyama2015}

\newblock \bibinfo{author}{Kiyama, H.}, \bibinfo{author}{Nakajima, T.},
  \bibinfo{author}{Teraoka, S.}, \bibinfo{author}{Oiwa, A.} \&
  \bibinfo{author}{Tarucha, S.}
\newblock \emph{\bibinfo{journal}{Physical Review B}}
  \textbf{\bibinfo{volume}{91}}, \bibinfo{pages}{155302}
  (\bibinfo{year}{2015}).

\bibitem{Gharavi2016}

\newblock \bibinfo{author}{Gharavi, K.}, \bibinfo{author}{Hoving, D.} \&
  \bibinfo{author}{Baugh, J.}
\newblock \emph{\bibinfo{journal}{Physical Review B}}
  \textbf{\bibinfo{volume}{94}}, \bibinfo{pages}{155417}
  (\bibinfo{year}{2016}).

\bibitem{Deng2016}

\newblock \bibinfo{author}{Deng, M.~T.} \emph{et~al.}
\newblock \emph{\bibinfo{journal}{Science}} \textbf{\bibinfo{volume}{354}},
  \bibinfo{pages}{1557--1562} (\bibinfo{year}{2016}).

\end{thebibliography}

\end{document}